\documentclass{aa} 
\usepackage{amsmath}
\usepackage{amssymb}
\usepackage[varg]{txfonts}
\usepackage{natbib}
\usepackage{acro}[=v2] 
\usepackage{graphicx}
\usepackage{dcolumn}
\usepackage{amsmath}
\usepackage{bm}
\usepackage{float}
\usepackage{enumitem}
\usepackage[skip=8pt plus1pt, indent=10pt]{parskip}
\usepackage{lipsum}
\usepackage{xcolor}
\usepackage{hyperref}
\usepackage{footmisc}
\usepackage{changes}

\renewcommand{\deleted}[1]{}
\newcommand{\gc}{\text{de}}
\newcommand{\aCC}{\text{de}}
\newcommand{\rphys}{\text{phys}}
\newcommand{\sprim}{}
\newcommand{\sini}{\text{early}}
\newcommand{\newmodel}{$\Lambda$CDM extension}
\newcommand{\syidx}[1]{%
	\acs{#1}%
}

\DeclareAcronym{SYMed}{
	short=\rho,
	long=Energy density,
	class=symbol
}
\DeclareAcronym{SYMedbg}{
	short=\bar{\rho},
	long=Energy density of background universe,
	class=symbol
}
\DeclareAcronym{SYMedavg}{
	short=\langle\rho\rangle,
	long=Energy density averaged over oscillation period,
	class=symbol
}
\DeclareAcronym{SYMpr}{
	short=p,
	long=Pressure,
	class=symbol
}
\DeclareAcronym{SYMprbg}{
	short=\bar{p},
	long=Pressure of background universe,
	class=symbol
}
\DeclareAcronym{SYMpravg}{
	short=\langle p\rangle,
	long=Pressure averaged over oscillation period,
	class=symbol
}
\DeclareAcronym{SYMdenspar}{
	short=\Omega,
	long=Density parameter for component $i$,
	class=symbol
}
\DeclareAcronym{SYMDE}{
	short=\Lambda,
	long=Dark Energy,
	class=symbol
}
\DeclareAcronym{SYMH}{
	short=H,
	long=Hubble parameter in cosmic time,
	class=symbol
}
\DeclareAcronym{SYMHc}{
	short=\mathcal{H},
	long=Hubble parameter in conformal time,
	class=symbol
}
\DeclareAcronym{SYMG}{
	short=G,
	long=Gravitational constant,
	class=symbol
}
\DeclareAcronym{SYMc}{
	short=c,
	long=Speed of light,
	class=symbol
}
\DeclareAcronym{SYMa}{
	short=a,
	long=Scalefactor a(t),
	class=symbol
}
\DeclareAcronym{SYMg}{
	short=g,
	long=FLRW metric,
	class=symbol
}
\DeclareAcronym{SYMgbg}{
	short=\bar{g},
	long=FLRW metric of background,
	class=symbol
}
\DeclareAcronym{SYMLdens}{
	short=\mathcal{L},
	long=Lagrangian density,
	class=symbol
}
\DeclareAcronym{SYMsfbec}{
	short=\psi,
	long=wave function of BEC,
	class=symbol
}
\DeclareAcronym{SYMsf}{
	short=\varphi,
	long=Scalar Field,
	class=symbol
}
\DeclareAcronym{SYMsfperturb}{
	short=\phi,
	long=Perturbations of Scalar Field,
	class=symbol
}
\DeclareAcronym{SYMvpot}{
	short=V,
	long=Potential of the Scalar Field Dark Matter,
	class=symbol
}
\DeclareAcronym{SYMlambdasi}{
	short=\lambda,
	long=Strength of self-interaction,
	class=symbol
}
\DeclareAcronym{SYMdenscontrast}{
	short=\delta,
	long=The density contrast describes the relative deviation of density from the average density of the universe,
	class=symbol
}
\DeclareAcronym{SYMcurvpar}{
	short=\kappa,
	long={curvature parameter: -1..open universe, +1..closed universe, 0..flat universe},
	class=symbol
}
\DeclareAcronym{SYMeos}{
	short=w,
	long={The equation of state (EOS) relates pressure to energy density},
	class=symbol
}
\DeclareAcronym{SYMderivD}{
	short=D,
	long={The covariant derivative with respect to $\mu$},
	class=symbol
}
\DeclareAcronym{SYMkron}{
	short=\delta_{ij},
	long={Kronecker delta},
	class=symbol
}
\DeclareAcronym{SYMmpsynh}{
	short=h,
	long={Metric perturbation in synchronous gauge},
	class=symbol
}
\DeclareAcronym{SYMmpsyneta}{
	short=\eta,
	long={Metric perturbation in synchronous gauge},
	class=symbol
}
\DeclareAcronym{SYMmpnewpot}{
	short=\Phi,
	long={Metric perturbation in Newtonian gauge -- potential},
	class=symbol
}
\DeclareAcronym{SYMmpnewlapse}{
	short=\Psi,
	long={Metric perturbation in Newtonian gauge -- laps function},
	class=symbol
}
\DeclareAcronym{SYMct}{
	short=\tau,
	long={Conformal time},
	class=symbol
}
\DeclareAcronym{SYMt}{
	short=t,
	long={Cosmological time},
	class=symbol
}
\DeclareAcronym{SYMemt}{
	short=T,
	long={Energy momentum tensor},
	class=symbol
}
\DeclareAcronym{SYMemtbg}{
	short=\bar{T},
	long={Energy momentum tensor for background},
	class=symbol
}
\DeclareAcronym{SYMemtvelocity}{
	short=\theta,
	long={Velocity divergence of the energy-momentum tensor},
	class=symbol
}
\DeclareAcronym{SYMemtstress}{
	short=\sigma,
	long={Anisotropic stress of the energy-momentum tensor},
	class=symbol
}
\DeclareAcronym{SYMradius}{
	short=R,
	long={Radius of the Universe},
	class=symbol
}
\usepackage{empheq}
\newcommand{\acidx}[1]{%
	\ac{#1}%
}
\DeclareAcronym{FLRW}{
	short=FLRW,
	long=Friedmann-Lema\^{i}tre-Robertson-Walker,
	class=acronym
}
\DeclareAcronym{DM}{
	short=DM,
	long=dark matter,
	class=acronym
}
\DeclareAcronym{DE}{
	short=DE,
	long=dark energy,
	class=acronym
}
\DeclareAcronym{CDM}{
	short=CDM,
	long=cold dark matter,
	class=acronym
}
\DeclareAcronym{HDM}{
	short=HDM,
	long=hot dark matter,
	class=acronym
}
\DeclareAcronym{WDM}{
	short=WDM,
	long=warm dark matter,
	class=acronym
}
\DeclareAcronym{BAOs}{
	short=BAOs,
	long=baryonic acoustic oscillations,
	class=acronym
}
\DeclareAcronym{LCDM}{
	short=$\Lambda$CDM,
	long=$\Lambda$ cold dark matter,
	class=acronym
}
\DeclareAcronym{wCDM}{
	short=wCDM,
	long=advanced $\Lambda$ cold dark matter,
	class=acronym
}
\DeclareAcronym{LSFDM}{
	short=$\Lambda$SFDM,
	long=$\Lambda$ scalar field dark matter,
	class=acronym
}
\DeclareAcronym{sCDM}{
	short=sCDM,
	long=standard cold dark matter,
	class=acronym
}
\DeclareAcronym{SFDM}{
	short=SFDM,
	long=scalar field dark matter,
	class=acronym
}
\DeclareAcronym{SF}{
	short=SF,
	long=scalar field,
	class=acronym
}
\DeclareAcronym{BEC}{
	short=BEC,
	long=Bose Einstein condensate,
	class=acronym
}
\DeclareAcronym{CMB}{
	short=CMB,
	long=cosmic microwave background,
	class=acronym
}
\DeclareAcronym{GR}{
	short=GR,
	long=general relativity,
	class=acronym
}
\DeclareAcronym{SR}{
	short=SR,
	long=special relativity,
	class=acronym
}
\DeclareAcronym{SUSY}{
	short=SUSY,
	long=super symmetry,
	class=acronym
}
\DeclareAcronym{SM}{
	short=SM,
	long=standard model of particle physics,
	class=acronym
}
\DeclareAcronym{BBN}{
	short=BBN,
	long=big bang nucleosynthesis,
	class=acronym
}
\DeclareAcronym{WIMP}{
	short=WIMP,
	long=weakly interacting massive particle,
	class=acronym
}
\DeclareAcronym{SDSS}{
	short=SDSS,
	long=Sloan Digital Sky Survey,
	class=acronym
}
\DeclareAcronym{SN1a}{
	short=SN Ia,
	long=supernova(e) Type Ia,
	class=acronym
}
\DeclareAcronym{ALPs}{
	short=ALPs,
	long=axion-like particles,
	class=acronym
}
\DeclareAcronym{ULAs}{
	short=ULAs,
	long=ultra-light axions,
	class=acronym
}
\DeclareAcronym{FDM}{
	short=FDM,
	long=fuzzy cold dark matter,
	class=acronym
}
\DeclareAcronym{BECDM}{
	short=BECDM,
	long=Bose-Einstein-condesate dark matter,
	class=acronym
}
\DeclareAcronym{QCD}{
	short=QCD,
	long=quantum chromo dynamics,
	class=acronym
}
\DeclareAcronym{SI}{
	short=SI,
	long=self-interaction,
	class=acronym
}
\DeclareAcronym{CC}{
	short=CC,
	long=cosmological constant,
	class=acronym
}
\DeclareAcronym{EOS}{
	short=EoS,
	long=equation of state,
	class=acronym
}
\DeclareAcronym{EOM}{
	short=EoM,
	long=equation of motion,
	class=acronym
}
\DeclareAcronym{KGE}{
	short=KGE,
	long=Klein-Gordon equation,
	class=acronym
}
\DeclareAcronym{IC}{
	short=ICs,
	long=initial conditions,
	class=acronym
}
\DeclareAcronym{CLASS}{
	short=CLASS,
	long=Cosmic Linear Anisotropy Solving System,
	class=acronym
}
\DeclareAcronym{IVP}{
	short=IVP,
	long=initial value problem,
	class=acronym
}
\DeclareAcronym{QFT}{
	short=QFT,
	long=quantum field theory,
	class=acronym
}
\DeclareAcronym{CFS}{
	short=CFS,
	long=causal fermion systems,
	class=acronym
}
\DeclareAcronym{EdS}{
	short=EdS,
	long=Einstein-de Sitter,
	class=acronym
}
\newcommand{\glo}[1]{%
	{\acs{#1}}%
}
\newcommand{\emidx}[2][]
{%
	\ifx&#1&%
	``#2''%
	\index{#2}%
	\else%
	#1#2#1%
	\index{#2}%
	\fi%
}

\newcommand{\TBDgl}{%
	\colorbox{yellow!30}{%
		\parbox{\dimexpr\linewidth-2\fboxsep\relax}{%
			Glossary to be done
		}%
	}
}
\DeclareAcronym{staticuniverse}
{
	short=static universe,
	long=\textit{Is a universe, that does not expand},
	extra=\\\TBDgl,
	class=glossary
}
\DeclareAcronym{flrwuniverse}
{
	short=\ac{FLRW} universe,
	long=\textit{Is a universe that is described by the \glo{flrwmetric}},
	extra=\\\TBDgl,
	class=glossary
}

\DeclareAcronym{flrwmetric}
{
	short=\ac{FLRW} metric,
	long=\textit{This metric is the most general type of a metric describing a homogeneous and isotropic spacetime},
	extra=\\\TBDgl,
	class=glossary
}

\DeclareAcronym{cosmoprinciple}
{
	short=cosmological principle,
	long=\textit{The cosmological principle refers to a universe, which is homogeneous (all locations in space are equivalent) and isotropic (all directions in space are equivalent)},
	extra=\\\TBDgl,
	class=glossary
}

\DeclareAcronym{LCDMmodel}
{
	short=$\Lambda$CDM model,
	long=\textit{Is a cosmological model, belonging to the family of the \glo{flrwuniverse} and has \ac{CDM} and \ac{DE}},
	extra=\\\TBDgl,
	class=glossary
}

\DeclareAcronym{LSFDMmodel}
{
	short=$\Lambda$SFDM model,
	long=\textit{Is a cosmological model based on the \glo{LCDMmodel}, where the \acidx{CDM} component is \acidx{SFDM} instead of the standard \acidx{WIMP} particles},
	extra=\\\TBDgl,
	class=glossary
}

\DeclareAcronym{friedmannequations}
{
	short=Friedmann equations,
	long=\textit{Friedmann derived a set of equations, that describe the evolution of the universe under the assumption of the \glo{cosmoprinciple}},
	extra=\\\TBDgl,
	class=glossary
}

\DeclareAcronym{friedmannequation}
{
	short=Friedmann equation,
	long=\textit{The term Friedmann equation refers to the first of the two Friedmann equations},
	class=glossary
}

\DeclareAcronym{decelerationequation}
{
	short=deceleration equation,
	long=\textit{The deceleration equation refers to the second of the two Friedmann equations},
	class=glossary
}

\DeclareAcronym{scalarfield}
{
	short=scalar field,
	long=\textit{The scalar field describes bosonic matter, that behaves like a \ac{BEC}},
	extra=\\\TBDgl,
	class=glossary
}

\DeclareAcronym{inflaton}
{
	short=Inflaton,
	long=\textit{The Inflaton is a \glo{scalarfield} that caused the inflationary epoch of the universe},
	extra=\\\TBDgl,
	class=glossary
}
\DeclareAcronym{inflation}
{
	short=inflation,
	long=\textit{The inflation is a very short period in time in the very early universe, where the universe expanded by many orders of magnitude},
	extra=\\\TBDgl,
	class=glossary
}
\DeclareAcronym{inflationaryphase}
{
	short=inflationary phase,
	long=\textit{The inflationary phase denotes the era of inflation in the evolution of the universe. see \glo{inflation}},
	extra=\\\TBDgl,
	class=glossary
}
\DeclareAcronym{cosmologicalconstant}
{
	short=cosmological constant,
	long=\textit{The cosmological constant is denoted by the symbol $\Lambda$ and is responsible for the accelerated expansion in the current epoch of the universe},
	extra=\\\TBDgl,
	class=glossary
}

\DeclareAcronym{criticaldensity}
{
	short=critical density,
	long=\textit{A universe with its density as big as the critical density will expand forever, but expansion stops at infinity. This universe is also called to be flat},
	extra=\\\TBDgl,
	class=glossary
}

\DeclareAcronym{densityparameter}
{
	short=density parameter,
	long=\textit{The density parameter...},
	extra=\\\TBDgl,
	class=glossary
}

\DeclareAcronym{lambdadominateduniverse}
{
	short=$\Lambda$~dominated universe,
	long=\textit{This term denotes our universe in the current epoch, where the \glo{darkenergy} dominates over all other constituents of the universe},
	extra=\\\TBDgl,
	class=glossary
}

\DeclareAcronym{radiationdominateduniverse}
{
	short=radiation dominated universe,
	long=\textit{In the early universe, its energy content was dominated by radiation (photons and relativistic neutrinos). The expansion of the universe diluted radiation more intensively than matter. Therefore, this era came to an end and the universe was then dominated by matter},
	extra=\\\TBDgl,
	class=glossary
}

\DeclareAcronym{matterdominateduniverse}
{
	short=matter dominated universe,
	long=\textit{After radiation has been diluted, matter (baryonic matter and Dark Matter) became the dominating energy in the universe. This era is lasting until today},
	extra=\\\TBDgl,
	class=glossary
}

\DeclareAcronym{darkenergy}
{
	short=dark energy,
	long=\textit{Dark Energy is a type of energy, which was discovered in the late 1990's and which is thought to be the reason for the accelerated expansion of the universe we see today},
	extra=\\\TBDgl,
	class=glossary
}

\DeclareAcronym{densitycontrast}
{
	short=density contrast,
	long=\textit{The density contrast is a quantity, that measures the relative deviation of the density at a point in space from the average density of the universe at a specific point in time. Mostly, it is denoted by the symbol $\delta$},
	extra=\\\TBDgl,
	class=glossary
}

\DeclareAcronym{flatnessproblem}
{
	short=flatness problem,
	long=\textit{The flatness problem is a problem in cosmology},
	extra=\\\TBDgl,
	class=glossary
}

\DeclareAcronym{horizonproblem}
{
	short=horizon problem,
	long=\textit{The horizon problem is a problem in cosmology},
	extra=\\\TBDgl,
	class=glossary
}

\DeclareAcronym{primordialpowerspectrum}
{
	short=primordial power spectrum,
	long=\textit{The primordial power spectrum is a description of the density perturbations at the end of \glo{inflation}},
	extra=\\\TBDgl,
	class=glossary
}

\DeclareAcronym{powerspectrum}
{
	short=power spectrum,
	long=\textit{The power spectrum is ...},
	extra=\\\TBDgl,
	class=glossary
}

\DeclareAcronym{harrisonzeldovichspectrum}
{
	short=Harrison Zel’dovich spectrum,
	long=\textit{The Harrison Zel’dovich, developed by \textsc{Harrison} and \textsc{Zel’dovich} is a theoretical description of a power spectrum with a power law},
	extra=\\\TBDgl,
	class=glossary
}

\DeclareAcronym{linearstructureformation}
{
	short=linear structure formation,
	long=\textit{Linear structure formation denotes a regime of structure formation, where the \glo{densitycontrast} is smaller than unity, such that all equations used to model the corresponding processes can be of linear order},
	extra=\\\TBDgl,
	class=glossary
}

\DeclareAcronym{cosmicmicrowavebackground}
{
	short=cosmic microwave background,
	long=\textit{The cosmic microwave background is the relic radiation of the Big Bang},
	extra=\\\TBDgl,
	class=glossary
}

\DeclareAcronym{virialradius}
{
	short=virial radius,
	long=\textit{The virial radius defines the size of an astrophysical object, that is in dynamical equilibrium. On the basis of theoretical calculations, the \glo{densitycontrast} is defined to be $200$},
	extra=\\\TBDgl,
	class=glossary
}

\DeclareAcronym{virialmass}
{
	short=virial mass,
	long=\textit{The virial mass defines the size of an astrophysical object, that is in dynamical equilibrium. On the basis of theoretical calculations, the \glo{densitycontrast} is defined to be $200$},
	extra=\\\TBDgl,
	class=glossary
}

\DeclareAcronym{hubbleflow}
{
	short=Hubble flow,
	long=\textit{The notion "Hubble flow" describes the situation, where matter "follows" the expansion of the universe},
	extra=\\\TBDgl,
	class=glossary
}

\DeclareAcronym{hubbleparameter}
{
	short=Hubble parameter,
	long=\textit{The notion "Hubble parameter" describes ...},
	extra=\\\TBDgl,
	class=glossary
}

\DeclareAcronym{hubblesphere}
{
	short=Hubble sphere,
	long=\textit{The notion "Hubble sphere" describes ...},
	extra=\\\TBDgl,
	class=glossary
}

\DeclareAcronym{hubbleconstant}
{
	short=Hubble constant,
	long=\textit{When \textsc{Hubble} discovered the \glo{hubbleslaw}, he found that the relation between the distance of a galaxy and the velocity the galaxy moves away from us, is defined by a constant. This constant was then called the Hubble constant. Meanwhile it is known that this quantity is not a constant, but varies with time. Therefore, it is now called generally \glo{hubbleparameter}.},
	extra=\\\TBDgl,
	class=glossary
}

\DeclareAcronym{hubbleslaw}
{
	short=Hubble's law,
	long=\textit{When \textsc{Hubble}, for the first time, observed the expansion of the universe, he found that galaxies move faster away from us the farther they are from us. This relation is called Hubble's law},
	extra=\\\TBDgl,
	class=glossary
}

\DeclareAcronym{conformaltime}
{
	short=conformal time,
	long=\textit{The notion "Conformal time" describes ...},
	extra=\\\TBDgl,
	class=glossary
}

\DeclareAcronym{MILNEuniverse}
{
	short=Milne universe,
	long=\textit{The Milne universe is a cosmological model developed by \textsc{Edward Arthur Milne} in 1933, where it was the first time to formulate the \glo{cosmoprinciple}},
	extra=\\\TBDgl,
	class=glossary
}

\DeclareAcronym{cosmoterm}
{
	short=cosmological term,
	long=\textit{\textsc{Einstein} added this term to his equations of \acidx{GR} to guarantee a \glo{staticuniverse}, which includes the \glo{cosmoconstant}},
	extra=\\\TBDgl,
	class=glossary
}

\DeclareAcronym{cosmoconstant}
{
	short=cosmological constant,
	long=\textit{\textsc{Einstein} added the \glo{cosmoterm} to his equations of \acidx{GR} to guarantee a \glo{staticuniverse}, he used this constant to generate an repulsive force to avoid the contraction of the universe caused by gravity. He used the symbol $\Lambda$ for this constant, which is nowadays used to denote \acidx{DE}},
	extra=\\\TBDgl,
	class=glossary
}

\DeclareAcronym{bigbang}
{
	short=big bang,
	long=\textit{The Big Bang was introduced by \textsc{Lema\^{i}tre} as he realized the expansion of the universe and concluded that the universe formed in a single event out of a "primeval atom"},
	extra=\\\TBDgl,
	class=glossary
}

\DeclareAcronym{hotbigbangmodel}
{
	short=hot Bing Bang model,
	long=\textit{This model states that the universe came into existence in an \glo{bigbang} event creating a extremely hot and dense environment. The subsequent evolution is determined by the laws of thermodynamics, where due to the expansion the universe cooled and diluted.},
	extra=\\\TBDgl,
	class=glossary
}

\DeclareAcronym{virialtheorem}
{
	short=virial theorem,
	long=\textit{The virial theorem describes the ratio of different types of energy in a system, being in dynamical equilibrium},
	extra=\\\TBDgl,
	class=glossary
}

\DeclareAcronym{darkmatter}
{
	short=Dark Matter,
	long=\textit{When \textsc{Fritz Zwicky} investigated the motions of the galaxies in the Coma cluster, he found that there is not enough matter to gravitationally bind the galaxies to the cluster. He concluded, that there is a type of matter that we cannot see. So he called it Dark Matter},
	extra=\\\TBDgl,
	class=glossary
}

\DeclareAcronym{cosmicweb}
{
	short=cosmic web,
	long=\textit{The large scale structure of the universe, consisting...},
	extra=\\\TBDgl,
	class=glossary
}

\DeclareAcronym{weakforce}
{
	short=weak force,
	long=\textit{The weak force...},
	extra=\\\TBDgl,
	class=glossary
}

\DeclareAcronym{axion}
{
	short=axions,
	long=\textit{The axions...},
	extra=\\\TBDgl,
	class=glossary
}

\DeclareAcronym{sterileneutrinos}
{
	short=sterile neutrinos,
	long=\textit{The sterile neutrinos...},
	extra=\\\TBDgl,
	class=glossary
}

\DeclareAcronym{selfinteractingDM}
{
	short=self-interacting DM,
	long=\textit{The self-interacting DM...},
	extra=\\\TBDgl,
	class=glossary
}

\DeclareAcronym{SMcosmo}
{
	short=Standard Model of Cosmology,
	long=\textit{The Standard Model of Cosmology is described in section \ref{sec:SMcosmo}},
	class=glossary
}

\DeclareAcronym{CMcosmo}
{
	short=Concordance Model of Cosmology,
	long=\textit{see \glo{SMcosmo}},
	class=glossary
}

\DeclareAcronym{debrogliewavelength}
{
	short=de Broglie wavelength,
	long=\textit{The de Broglie wavelength...},
	extra=\\\TBDgl,
	class=glossary
}

\DeclareAcronym{missingsatelitesproblem}
{
	short=missing-satellites problem,
	long=\textit{The missing-satellites problem...},
	extra=\\\TBDgl,
	class=glossary
}

\DeclareAcronym{cuspcoreproblem}
{
	short=cusp-core problem,
	long=\textit{The cusp-core problem...},
	extra=\\\TBDgl,
	class=glossary
}

\DeclareAcronym{toobigtofailproblem}
{
	short=too-big-to-fail problem,
	long=\textit{The too-big-to-fail problem...},
	extra=\\\TBDgl,
	class=glossary
}

\DeclareAcronym{missingmagneticmonopoles}
{
	short=missing magnetic monopoles,
	long=\textit{The missing magnetic monopoles problem...},
	extra=\\\TBDgl,
	class=glossary
}

\DeclareAcronym{strongcpproblem}
{
	short=strong CP problem,
	long=\textit{The strong CP problem...},
	extra=\\\TBDgl,
	class=glossary
}

\DeclareAcronym{thomasfermiregime}
{
	short=Thomas-Fermi regime,
	long=\textit{Thomas-Fermi regime...},
	extra=\\\TBDgl,
	class=glossary
}

\DeclareAcronym{kleingordonequation}
{
	short=Klein-Gordon equation,
	long=\textit{Klein-Gordon equation...},
	extra=\\\TBDgl,
	class=glossary
}

\DeclareAcronym{bulletcluster}
{
	short=Bullet Cluster,
	long=\textit{The Bullet Cluster...},
	extra=\\\TBDgl,
	class=glossary
}

\DeclareAcronym{QCDaxion}
{
	short=QCD axion,
	long=\textit{Axion particle proposed to solve strong CP problem in QCD},
	extra=\\\TBDgl,
	class=glossary
}
\DeclareAcronym{wCDMmodel}
{
	short=wCDM model,
	long=\textit{Is a cosmological model, extending the $\Lambda$CDM model to integrate DM halo collapse and virialisation and providing an explanation of the physical nature of Dark Energy},
	extra=\\\TBDgl,
	class=glossary
}
\DeclareAcronym{owCDMmodel}
{
	short=owCDM model,
	long=\textit{Is a cosmological model, extending the $\Lambda$CDM model to integrate DM halo collapse and virialisation and providing an explanation of the physical nature of Dark Energy},
	extra=\\\TBDgl,
	class=glossary
}
\definecolor{MYCOLOR}{RGB}{255,0,0}
\definecolor{mycolor}{RGB}{255,0,0}

\newcommand{\tb}[1]{\textbf{#1}}

\newcommand{\q}[1]{``#1''}
\newcommand{\qc}[2]{``#1#2''}

\newcommand{\qemph}[1]{\textit{#1}}

\newcommand{\CCLambda}[1]{$\syidx{SYMDE}$}
\newcommand{\Hexprate}[1]{$H$}
\newcommand{\Hconst}[1]{$H_0$}

\makeatletter
\newcommand*{\hyperlinkcite}[1]{\hyper@link{cite}{cite.#1}}
\makeatother

\begin{document}
\title{A \boldmath$\Lambda$CDM Extension Explaining the Hubble Tension and the Spatial Curvature $\Omega_{k,0} = - 0.012 \pm 0.010$ Measured by the Final PR4 of the Planck Mission}
\titlerunning{A \boldmath$\Lambda$CDM Extension Explaining the Hubble Tension and the Spatial Curvature}
%
%
\date{\today}
\author{Horst Foidl\inst{1,2}\thanks{horst.foidl@outlook.com}
	\and
	Tanja Rindler-Daller\inst{1,2,3}\thanks{tanja.rindler-daller@univie.ac.at}
}

\institute{Institut f\"ur Astrophysik, Universit\"atssternwarte Wien,
	Fakult\"at f\"ur Geowissenschaften, Geographie und Astronomie,\\
	Universit\"at Wien, T\"urkenschanzstr.17, A-1180 Vienna, Austria
	\and
	Vienna International School of Earth and Space Sciences, Universit\"at Wien, Josef-Holaubek-Platz 2,
	A-1090 Vienna, Austria
	\and
	Wolfgang Pauli Institut, Oskar-Morgenstern-Platz 1, A-1090 Vienna, Austria
}

\date{\today}
%
\abstract{
The measurements of the CMB have determined the cosmological parameters with high accuracy, and the observation of the flatness of space have contributed to the status of the concordance $\Lambda$CDM model. However, the cosmological constant $\Lambda$, necessary to close the model to critical density, remains an open conundrum. We explore the observed late-time accelerated expansion of the Universe, where we consider that the Friedmann equation describes the expansion history of FLRW universes in the local reference frame of freely falling comoving observers, which perceive flat, homogeneous and isotropic space in their local inertial system, where, as a consequence of the equivalence principle, special relativity applies. We use this fact to propose an extension to $\Lambda$CDM, incorporating the initial conditions of the background universe, comprising the initial energy densities as well as the initial post big bang expansion rate. The observed late-time accelerated expansion is then attributed to a kinematic effect akin to a dark energy component.  Choosing the same $\Omega_{m,0} \simeq 0.3$ as $\Lambda$CDM, its equation of state $w_{de} \simeq -0.8$. Furthermore, we include the impact on the expansion history caused by the cosmic web of the late Universe, once voids dominate its volume, and find that the initially constant $w_{de}$ becomes time-dependent, evolving to a value of $w_{de} \simeq -0.9$ at the present. While this impact by voids is minor, it is sufficient to provide a solution to the Hubble tension problem. We use CLASS to calculate the expansion history and power spectra of our extension and compare our results to concordance $\Lambda$CDM and to observations. We find that our model agrees well with current data, in particular with the final data release PR4 of the Planck mission, where it explains the reported spatial curvature of $\Omega_{k,0} = - 0.012 \pm 0.010$.
}
\maketitle
\nolinenumbers
%
\section{Introduction}\label{sec:introductionDE}
%
\acresetall
The nature of the cosmological constant $\Lambda$ has been a challenging open problem in modern cosmology. The issue has intensified over the years, as the current concordance model of $\Lambda$CDM -- with \acidx{CDM} the other open conundrum \mbox{--,} continues to pass many observational tests, especially on the large scales of the cosmic web, or the \acidx{CMB} radiation. In particular, the \acidx{CMB} measurements have invigorated the flat universe interpretation, i.e. that the space of the background universe has no curvature.

However, it was reasoned that the amount of baryonic matter predicted by big bang nucleosynthesis was far too small to explain a flat universe (e.g., \citet{Steigman2007}), and even the addition of a substantial amount of \acidx{DM} in the form of \acidx{CDM}, whose presence was not in doubt any longer by the late 1980s, was not sufficient to get a universe at critical density, in order to get a flat geometry of space. In fact, age determinations of the oldest known stars ruled out the \acidx{EdS} model as a viable model, which could have otherwise been a model for a CDM-dominated universe at the present, with its critical density provided by matter only (see e.g. \citet{Weinberg2008}). Finally, in accordance with the predictions from inflation (\citet{Guth1981}), the observations of the \acidx{CMB} by the balloon-based BOOMERanG experiment (\citet{Bernardis2000} and \citet{MacTavish2006})), as well as observations with increasing accuracy by the space missions COBE (\citet{Smoot1992}), WMAP (\citet{Hinshaw2013}) and Planck (\citet{Collaboration2020}), boosted the flat universe interpretation. This, and the previous discovery of the accelerated expansion of the Universe (\citet{Perlmutter1999,Schmidt1998,Riess1998,Perlmutter2003}), led to the empirical addition of the cosmological constant $\Lambda$, in order to get a universe at critical density, that is a \qemph{flat universe} and the \acuse{LCDM}\acidx{LCDM} concordance model became the standard model of cosmology.

Despite the great success of \acidx{LCDM}, the still unknown nature of \acidx{CDM} and the nature of the cosmological constant $\Lambda$, which, according to measurements, each contribute roughly 25\% and 70\%, respectively, to the present-day energy density of the Universe, remain pending questions. Many cosmological observation campaigns analyze their data, not only in light of testing \acidx{LCDM}, but also to examine possible extensions to \acidx{LCDM}, which replace the cosmological constant $\Lambda$ with different alternative models of \acidx{DE}, which employ a time-dependent \acidx{EOS} parameter $w = p/\rho$ (The energy density $\rho$ and the pressure $p$ are commonly understood as components of some physically or operationally defined DE component).

Prominent examples of campaigns include the Dark Energy Survey, see for example, the Year 3 (DES-Y3) results in \citet{Abbott2022a,Abbott2023}, or the CMB measurements by \citet{Collaboration2020}. They use the so-called \q{CPL parametrization} by \citet{Chevallier2001,Linder2003} for the \acidx{EOS} parameter,
%
\begin{equation} \label{eq:EQCPLH0}%
	w(a) = w_0 + \left(1 - a \right) w_a,
\end{equation}
which is defined for $a \in [0,1]$, where $a$ is the scale factor\footnote{In practice, cosmological codes, including the one we use (CLASS), in the default configuration compute observables not earlier than at $a=10^{-14}$.}. The CPL parametrization is simple and empirically defined, with no physical motivation for the evolution of $w(a)$ being linear in $a$. In \citet{Foidl2024} we exemplify, based on empirical arguments, that a cosmological model including a CPL-based \acidx{DE} component with an \acidx{EOS} parameter $w_{de}$ evolving from $-0.8$ to $-0.9$ describes the \acidx{CMB} temperature spectrum of \acidx{LCDM} and yields a Hubble constant $H_0$ being compatible with direct measurements (e.g., \citet{Riess2022}) in the local Universe.  In contrast, however, in this paper we propose physically motivated extensions to \acidx{LCDM} -- wCDM and owCDM -- with a time-dependent \acidx{EOS} parameter $w_{de}$ of its effective DE component also evolving from $-0.8$ to $-0.9$, which we would like to put up for discussion\footnote{We use these acronyms for our models, as they are used frequently in the literature to express an extension to $\Lambda$CDM, particularly for dynamical models of DE, without and with spatial curvature, respectively (see e.g. Dark Energy Survey Year 3 Results \citet{Abbott2023}, \citet{Collaboration2020} and the Particle Data Group Review, Chapter 28 \q{Dark Energy} \cite{PDG202223}).}.

In the first part of this paper, we reassess \acidx{LCDM}'s flat universe interpretation, where we consider that the \glo{friedmannequation} describes the expansion history of \glo{flrwuniverse}s in the local reference frame of comoving observers, which perceive flat, homogeneous and isotropic space in their local inertial system, based on the equivalence principle (see e.g., \citet{Weinberg1972,Weinberg2008}, \citet{Peacock1999} or \citet{Fliessbach2016}). Thus, the measurements of the \acidx{CMB}, reporting the flatness of space, confirm that we fulfill the criterion of a comoving observer to a high degree. We interpret the expansion history of the background universe as an \acidx{IVP}, where the post \glo{bigbang} \acidx{IC} in the early Universe are given by the initial densities of the cosmic components in the Universe (radiation, baryons and \acidx{DM}) and the post big bang expansion rate $\syidx{SYMH}_{\text{ini}}$.

In the second part of this paper, we find that the kinematic effects induced by the initial expansion rate (post big bang) and the initial density of the background universe, can be described in a straightforward manner by associating the operationally defined quantity $\rho_{\aCC}$, dark energy,
replacing $\rho_{\Lambda}$, with an \acidx{EOS}, being different from that of a cosmological constant. 
Choosing for the other parameters the same values as for the \acidx{LCDM} concordance model, our wCDM model yields $w_{\aCC} \sim -0.8$, which results in an expansion history very close to \acidx{LCDM} with its $w_{\Lambda} = -1$.

In the third part of the paper, we include in our \newmodel{} the effects of the cosmic voids, which dominate the volume of the Universe in the late stages of the evolution of the cosmic web. The resulting backreaction from voids also affects the evolution of the expansion rate. We consistently include this effect into $\rho_{\aCC}$, using void models informed by cosmological $\Lambda$CDM simulations of the previous literature, and parameterize it by deriving the necessary modifications to the heretofore constant \acidx{EOS} parameter $w_{\aCC}$, that becomes a function of time, or scale factor, respectively, once the voids dominate the volume of the Universe. In fact, the initial value of $w_{\aCC}(a)  \sim -0.8$ decreases to $w_{\aCC}(1) \sim -0.9$ at the present where the scale factor $a=1$. Thus, it gets closer to the \acidx{EOS} of a cosmological constant. As a result, the present-day value of the expansion rate $H_0$, i.e. the Hubble constant, shifts to a higher value, compared to the value it would have, if the EoS parameter remained constant throughout cosmic time (see also \citet{Foidl2024}).  Although the expansion rate is changed by only about $8\%$ due to the impact of voids, it is enough to provide a solution to the Hubble tension problem. 

By comparing our results with cosmological observations, we find that our \newmodel{} is in agreement with current data. To this end, we apply our proposed procedure by \citet{Foidl2024} to our model and fit it to the \acidx{LCDM} CMB spectrum and $H_0 = 73.04$~km/s/Mpc as determined by \citet{Riess2022} in the local Universe, where we, in addition to \acuse{wCDM}\acidx{wCDM}, consider also spatial curvature. The resulting owCDM model perfectly agrees with the final results of the Planck mission (\citet{Tristram2024}), in particular with the reported spatial curvature of $\Omega_{k,0} = - 0.012 \pm 0.010$.

This paper is organized as follows. In Sec.~\ref{sec:standardFriedmann} we recapitulate the basic equations for the evolution of the background universe in \acidx{FLRW} models. Sec~\ref{sec:FlatUniverse} investigates the flat universe interpretation for the general case of \acidx{FLRW} universes, followed by a discussion of the \acidx{IC} of the background universe.
Sec.~\ref{sec:EnhancedFriedmann} proposes a \newmodel{}, such that the cosmological model incorporates the post big bang initial conditions of the early Universe.
Sec.~\ref{sec:largescale} contains our treatment and derivation of the impact of cosmic voids onto the expansion history. This impact is minor, but significant enough that it provides a possible resolution to the Hubble tension problem\deleted{, as shown subsequently}. 
In the following Sec.~\ref{sec:resultsiCDM}, we present the results of the numerical simulations of our \newmodel{} wCDM, based on our amended version of the code \acidx{CLASS}, where we also compare our results with the concordance \glo{LCDMmodel}, as well as with observations. In Sec.~\ref{sec:resultsowCDM} we present the results of our owCDM model and compare it to Planck's PR4 (\citet{Tristram2024}).
Finally, in Sec.~\ref{sec:iCDMsummarydiscussion}, we summarize the presented concepts, results and implications, also in light of cosmological observations.
%
\section{Basic Equations for the Expansion History in FLRW Models}\label{sec:standardFriedmann}

First, we recapitulate the equations involved in the description of the evolution of the homogeneous and isotropic background universe, we need in the following.
As gravity is the only force acting on cosmological length scales, it determines the evolution of the background universe and is described by Einsteins' field equations
%
\begin{equation} \label{eq:iCDMEQEinsteinNL}%
	R_{\mu\nu} - \frac{1}{2} g_{\mu\nu} R  = \frac{8 \pi G}{c^4} T_{\mu \nu},
\end{equation}
with the Ricci tensor $R_{\mu \nu}$ and the Ricci scalar $R$. 
The left-hand side (lhs) of the equation is often expressed as $E_{\mu \nu}$, the Einstein tensor. The right-hand side (rhs) contains the energy-momentum tensor $T_{\mu \nu}$, which includes the cosmic inventory of given cosmological models. The energy-momentum tensor $T_{\mu \nu}$ determines the curvature of space, expressed by the metric $g_{\mu \nu}$, which is also included in the Ricci tensor $R_{\mu \nu}$ and the Ricci scalar $R$.  Cosmological models differ in their assumptions on the nature and amount of the cosmic components encoded in $T_{\mu \nu}$. As such, all models are subject to observational constraints.

The geometry of a universe with constant curvature is described by applying the Riemannian formalism of curved surfaces and was developed by \citet{Robertson1936a} and \citet{Walker1937} based on  Milne's idea of a kinematically determined universe (\citet{Milne1933}) and preceding works by Friedmann \cite{Friedmann1922,Friedmann1924} and \citet{Lemaitre1927}. 
In spherical coordinates ($r, \theta, \phi$), the line element of the metric reads as
%
\begin{subequations} \label{eq:iCDMEQflrwbgk}%
	\begin{empheq}{align}
			ds^2 &= c^2 dt^2 - R^2(t) \left(\frac{dr^2}{1-k r^2} + r^2 d\Omega^2\right) \label{subeqn:EQflrwbgk1} \\
			d\Omega^2 &= d\theta^2 + \sin^2 \theta d \phi^2 \label{subeqn:EQflrwbgk3},
		\end{empheq}
\end{subequations}
with $k$ the curvature index and its values $+1$ (closed universe), $0$ (flat universe), $-1$ (open universe). The spatial coordinates ($r, \theta, \phi$) are defined in the comoving frame where $r$, $\theta$ and $\phi$ remain \q{fixed}, corresponding to the assumption of spaces of constant global curvature (the geometry of the Universe).
$R(t)$ is the \q{radius of curvature}, as called by \citet{Kolb1990}, at cosmic time $t$ (for $k=\pm 1$), with the dimension of length.

Applying the \acidx{FLRW} metric \eqref{eq:iCDMEQflrwbgk} to the metric tensor $g_{\mu \nu}$ in the Einstein equations \eqref{eq:iCDMEQEinsteinNL}, 
the energy-momentum tensor $T_{\mu \nu}$ then takes a perfect-fluid form, which reads 
%
\begin{equation} \label{eq:EQTperfectfluid}%
	T^{\mu\nu} = \left(\syidx{SYMed} + \frac{\syidx{SYMpr}}{\syidx{SYMc}^2}  \right) u^{\mu} u^{\nu} - g^{\mu\nu} \syidx{SYMpr}.
\end{equation}
with $u^{\mu}$ and $u^{\nu}$ the four velocities $\gamma(c, \vec{u})$ and \mbox{$\gamma=1/\sqrt{1-u^2/c^2}$}.
The time-time component of the solution to the Einstein equations \eqref{eq:iCDMEQEinsteinNL} yields the (first) \glo{friedmannequation} in the \qemph{classical version}, as derived by \citet{Friedmann1922}, see for example \citet{Kolb1990},
%
\begin{equation} \label{eq:EQfriedmannsgeneralNLK}%
	H^{2}(t) = \frac{8 \pi G}{3 c^{2}} \rho - \frac{k c^{2}}{a^2(t)},
\end{equation}
which describes the dynamics of the evolution of the background universe. Here, $\syidx{SYMed}$ refers to the entire energy density of the Universe; $k$ is the curvature index, defined in Eq.~\eqref{eq:iCDMEQflrwbgk} determining the geometry, as $k = +1$ for a closed (supercritical, density greater than the critical density) universe, $k = -1$ for an open (subcritical, density less than the critical density) universe and $k = 0$ for a flat geometry with critical density. In the forthcoming we use the notion \q{geometry} for the  curvature of the Universe as determined by the \acidx{FLRW} metric \eqref{eq:iCDMEQflrwbgk} and the \glo{friedmannequation}  \eqref{eq:EQfriedmannsgeneralNLK}, respectively. $a$ refers to the scale factor, with the dimension of length, which, owing to the symmetries imposed by the isotropy and homogeneity of the background universe, is a function of cosmic time $t$ and is defined as the relative \q{size} of an expanding or contracting universe, relative to its present-day size $\lvert a_0 \rvert = 1$. $H$ is the Hubble parameter
(we use the term expansion rate interchangeably) defined as
%
\begin{equation} \label{eq:iCDMEQH}%
	H(t) = \frac{\dot{a}}{a},
\end{equation}
where the dot refers to the derivative with respect to cosmic time $t$.
The space-space component yields the second \glo{friedmannequation}
\begin{equation} \label{eq:EQfriedmannsgeneralNLKII}%
	\frac{\ddot{\syidx{SYMa}}}{\syidx{SYMa}} = - \frac{4 \pi \syidx{SYMG}}{3 \syidx{SYMc}^{2}} \left(\syidx{SYMed} + 3\syidx{SYMpr}\right),
\end{equation}
which Friedmann called the deceleration equation. In recent literature it is called the acceleration equation.

Now, let us introduce the cosmic inventory which features the current concordance $\Lambda$CDM model. Also, we introduce some standard notions and equations which we need in the paper.
The energy densities of interest include cold dark matter (\q{CDM}), baryons (\q{b}) and radiation (\q{r}) (including photons and neutrinos). In the formulae, we also include the cosmological constant $\Lambda$, empirically added to the cosmic inventory to explain the flatness of space.

In order to study a variety of cosmological models, it has become customary to put \q{curvature} as well as the cosmological constant \q{$\Lambda$} into the energy-momentum tensor $T_{\mu \nu}$, by operationally defining \q{effective} energy densities for them, namely $\rho_{k} = -3 k c^2/(8\pi G a^2)$ for curvature (\q{k}), and $\rho_{\Lambda} = \Lambda c^2/(8\pi G)$ for the cosmological constant.
In the \glo{LCDMmodel}, $\rho_{k} = 0$ is prescribed, while $\rho_{\Lambda} \not= 0$, such that $\Lambda$ closes the Universe to critical density (see Eq.~\eqref{eq:EQclosureLCDMfull}). 

We stress that, although $\rho_k$ represents a geometric quantity, it has morphed into a \qc{substance}{,} or cosmic inventory,  described by $T_{\mu \nu}$, upon this standard operational procedure. Nevertheless, it is rightfully not regarded a physical constituent of the Universe, but simply a mathematical formalism, contributing an effective or artificial contribution to $T_{\mu \nu}$. On the other hand, $\Lambda$ is usually regarded as a real physical cosmic inventory in \acidx{LCDM}, which contributes to $T_{\mu \nu}$ basically in the same manner as matter and radiation.

The \glo{friedmannequation} in \qemph{modern} language reads as
%
\begin{equation} \label{eq:EQfriedmannLCDMnbg}%
	\begin{split}
		H^{2}(t) = \frac{8 \pi G}{3 c^{2}} \left[ \rho_{r}(t) + \rho_{b}(t) + \rho_{\text{CDM}}(t) + \rho_{k}(t) +		\rho_{\Lambda}(t) \right] 
	\end{split}
	{,}
\end{equation}
with the time-dependent background energy densities for radiation ($\syidx{SYMed}_{r}$), baryons ($\syidx{SYMed}_{b}$), CDM ($\syidx{SYMed}_{\text{CDM}}$), the curvature ($\syidx{SYMed}_{k}$), as well as the cosmological constant ($\rho_{\Lambda}$).
$H(t)$ is the \glo{hubbleparameter} and its present-day value\footnote{The literature has adopted the notational subscript \q{0} to denote present-day values, not the values at $t=0$.}, the Hubble constant, is denoted as $H_0$. The \glo{criticaldensity}, defining a flat universe, is given by
%
\begin{equation} \label{eq:iCDMEQcritdenst}%
	\rho_{\text{crit},t} = \frac{3 H^2(t) c^{2}}{8 \pi G},
\end{equation}
derived from \eqref{eq:EQfriedmannsgeneralNLK} with vanishing curvature term.
It is convenient to introduce the so-called density parameters or cosmological parameters
%
\begin{equation} \label{eq:iCDMEQdenspar}%
	\syidx{SYMdenspar}_{t,i} = \frac{\syidx{SYMed}_{i}(t)}{\syidx{SYMed}_{\text{crit},t}},
\end{equation}
where $i=$ CDM, b, r, etc., and
which are nothing but the background energy densities relative to the critical density \eqref{eq:iCDMEQcritdenst}. 

In order to solve the \glo{friedmannequation}, customarily, the energy conservation equation  is applied (for each component $i=$CDM, b, r,...), which reads 
%
\begin{equation} \label{eq:EQeconsnbg}%
	\frac{\partial \rho_i}{\partial t} + 3 H \left(\rho_i + p_i\right) = 0,
\end{equation}
where $\syidx{SYMed}_i$ and $\syidx{SYMpr}_i$ stand for the respective background energy densities and pressures\footnote{This equation assumes that there is no transformation between different components.}.
The energy densities and pressures are each related by their respective \acidx{EOS},
%
\begin{equation} \label{eq:EQeosnbg}%
	p_i(t) = w_i(t) \rho_i(t),
\end{equation}
where $w_i$ is often called the \acidx{EOS} parameter, which can also change with time, in general. However, in $\Lambda$CDM, $w_i$ is assumed to be a constant\footnote{However, for a detailed study of phase transitions in the early Universe, it is important to include e.g. a variable EoS of the radiation component, in order to take into account the reduction of relativistic degrees of freedom in the wake of the Universe's expansion.} for every component $i = $CDM, b, r, $\Lambda$ and k. 
Assuming a constant \acidx{EOS} parameter $w_i$ and using \eqref{eq:EQeosnbg} in \eqref{eq:EQeconsnbg}, it follows that $\dot{\rho_{i}}/\rho_{i} = -3(1+w_i)\dot{a}/a $, which is readily integrated to yield the well-known relation
%
\begin{equation} \label{eq:EQdensevolgen}%
	\rho_{i}(a) = \Omega_{i,0} \rho_{\text{crit},0} \: a^{-3(1+w_i)},
\end{equation}
which describes the evolution of the background energy densities as a function of the scale factor $a$ for constant $w_i$.
The background evolution of the standard cosmic components is thus given by
%
\begin{subequations} \label{eq:EQdensitiesLCDMfull}%
	\begin{empheq}{align}
		\rho_{r}(a) &= \Omega_{r,0} \rho_{\text{crit},0} / a^4  \label{subeqn:EQdensitiesLCDMfr}\\
		\rho_{m}(a) &= \Omega_{m,0} \rho_{\text{crit},0} / a^3  \label{subeqn:EQdensitiesLCDMfm}\\
		\rho_{k}(a) &= \Omega_{k,0} \rho_{\text{crit},0} / a^2  \label{subeqn:EQdensitiesLCDMfk}\\
		\rho_{\Lambda} &= \Omega_{\Lambda,0}\rho_{\text{crit},0}\label{subeqn:EQdensitiesLCDMfL},
	\end{empheq}
\end{subequations}
with \eqref{subeqn:EQdensitiesLCDMfr} for radiation (its \acidx{EOS} parameter in (\ref{eq:EQeosnbg}) is $w_r=1/3$); \eqref{subeqn:EQdensitiesLCDMfm} for baryonic matter and for CDM ($w_m=0$); \eqref{subeqn:EQdensitiesLCDMfk} for curvature ($w_k=-1/3$), and \eqref{subeqn:EQdensitiesLCDMfL} for the \glo{cosmoconstant} $\Lambda$ ($w_{\Lambda}=-1$).

The Friedmann equation can be alternatively written as an algebraic closure condition. At the present, it reads
%
\begin{equation} \label{eq:EQclosureLCDMfull}%
	1 = \Omega_{r,0} + \Omega_{b,0} + \Omega_{\text{CDM},0}  + \Omega_{k,0} + \Omega_{\Lambda,0}.
\end{equation}
In other words, Eq.~\eqref{eq:EQclosureLCDMfull} is the normalization of the Friedmann equation  \eqref{eq:EQfriedmannLCDMnbg} to the critical density.

\section{The Flat Universe Interpretation}\label{sec:FlatUniverse}

The \glo{LCDMmodel} is a member of the broader family of \acidx{FLRW} cosmological models. Furthermore, a flat space in the Universe is assumed on the grounds of the curvature of space, as measured for example by the observations of the \acidx{CMB} -- the flat universe interpretation, where the curvature term in the \glo{friedmannequation} \eqref{eq:EQfriedmannsgeneralNLK} is customarily interpreted to express the geometry of the Universe. We will now reassess this interpretation.

\subsection{Curvature in FLRW Universes}\label{sec:curvatureFLRW}

Let us elaborate on the curvature term appearing in the \glo{friedmannequation} \eqref{eq:EQfriedmannsgeneralNLK}, which is connected to the curvature factor in the \glo{flrwmetric} \eqref{eq:iCDMEQflrwbgk}.
The derivation of Eq.~\eqref{eq:EQfriedmannsgeneralNLK}, see for example \citet{Kolb1990}, suggests that the curvature term should not be confused with a contribution to the energy momentum tensor, which determines the Riemann tensor in the Einstein equations~\eqref{eq:iCDMEQEinsteinNL}. It is these equations which ought to determine the global curvature of space in the Universe.
We now reevaluate the interpretation of the curvature term of Eq.~\eqref{eq:EQfriedmannsgeneralNLK}, in order to discuss whether it describes the global curvature of space or not, as follows.

In a first step, we use the Einstein equations \eqref{eq:iCDMEQEinsteinNL} only, which describe the curvature of space, determined by the energy momentum tensor $\syidx{SYMemt}_{\mu\nu}$, that describes the distribution of energy (or matter) in the Universe. Presuming the \glo{cosmoprinciple}, $\syidx{SYMemt}_{\mu\nu}$ takes the form for a perfect fluid, given in Eq.~\eqref{eq:EQTperfectfluid}, where the respective densities and pressures of the cosmic components of interest are considered. Applying $\syidx{SYMemt}_{\mu\nu}$ and solving the Einstein equations \eqref{eq:iCDMEQEinsteinNL} yields the metric tensor $\syidx{SYMg}_{\mu\nu}$, which describes the global curvature of space in the Universe.
Considering, for example, the \acidx{EdS} universe, we apply $\syidx{SYMemt}_{\mu\nu}$ with the density at \glo{criticaldensity} and zero pressure for pressureless matter, which yields a non-flat metric tensor $\syidx{SYMg}_{\mu\nu}$. This is easily seen in a kind of cross-check by solving the \glo{friedmannequation} for the \acidx{EdS} universe $\syidx{SYMH}^2 = (8\pi \syidx{SYMG}/3c^2) \syidx{SYMed}_{\text{crit}}$, which yields $\syidx{SYMH}^2(\syidx{SYMa}) \propto \syidx{SYMa}^{-3}$. Thus, we see a deceleration of the expansion rate $\syidx{SYMH}$ in the course of the expansion of the \acidx{EdS} universe, which is caused by gravity (see e.g., \citet{Peacock1999}). It is important to note that no curvature term appears in Eq.~\eqref{eq:EQfriedmannsgeneralNLK} for the \acidx{EdS} universe.

In a second step, we follow the steps of the derivation of the \glo{friedmannequation}~\eqref{eq:EQfriedmannsgeneralNLK}, described in Sec.~\ref{sec:standardFriedmann} and reverse the procedure of step one. In contrast to the procedure of step one, determining the curvature of space for a given $\syidx{SYMemt}_{\mu\nu}$, we start by specifying the  metric tensor $\syidx{SYMg}_{\mu\nu}$ corresponding to the curvature term of Eq.~\eqref{eq:EQfriedmannsgeneralNLK}. Applying this metric tensor $\syidx{SYMg}_{\mu\nu}$ to the Einstein equations \eqref{eq:iCDMEQEinsteinNL} yields $\syidx{SYMemt}_{\mu\nu}$ of the corresponding energy densities\footnote{In general, this solution is not unique.}. The obtained energy momentum tensor(s) is(are) checked for agreement with the one we used in the first step.
We continue with the example of the \acidx{EdS} universe: the curvature term in \eqref{eq:EQfriedmannsgeneralNLK} vanishes for the \acidx{EdS} universe and the corresponding metric tensor $\syidx{SYMg}_{\mu\nu}$ describes flat space, which yields the energy momentum tensor $\syidx{SYMemt}_{\mu\nu}$ of empty space. This is different from a space at \glo{criticaldensity} in the \acidx{EdS} universe. Thus, in general, the curvature term does not express the spatial curvature\footnote{To falsify the assumption that the curvature term in Eq.~\eqref{eq:EQfriedmannsgeneralNLK} expresses the curvature of space, one contradicting example is sufficient.}. 
This is addressed by \citet{Robertson1936}, by using an \q{auxiliary} (mathematical) space associated with the  \glo{flrwmetric}~\eqref{eq:iCDMEQflrwbgk} (based on Riemannian geometry), which categorizes the dynamics of expansion of the background universe via the curvature index $k \in{\left\{-1,+1,0\right\}}$ as open (negative curvature), closed (positive curvature) and flat (no curvature), based on the energy density of the background universe relative to the \glo{criticaldensity}\footnote{The definition of $k$ refers to the density although the density does not appear in the metric. In the \acidx{FLRW} formalism this is addressed by the normalized Friedmann equation \eqref{eq:EQclosureLCDMfull}, relating density and curvature.  Consequently, in the \acidx{FLRW} formalism, the density parameters of subcritical and supercritical universes are also normalized to \glo{criticaldensity}, for example by considering the suitable amount of curvature. Remember, the curvature is not regarded as a physical contribution to the energy budget of the Universe.}. \citet{Walker1937} uses the term \q{Riemannian space} for the space connected to the metric. 

However, with regard to \acidx{LCDM}, the question arises as to how it is possible that observations of the \acidx{CMB} report the flatness of space given that the Universe is not empty, which we discuss as follows.
The \glo{friedmannequation} \eqref{eq:EQfriedmannsgeneralNLK} describes the dynamics of the expansion of the background universe, expressed by the evolution of the expansion rate $\syidx{SYMH}(t)$ in the local reference frame of observers comoving with the expansion, moving on geodesics, that is freely falling \acidx{FLRW} observers\footnote{The comoving observer is called fundamental observer by \citet{Robertson1935}. This term is also sometimes used in the literature.}. The expansion rate $\syidx{SYMH}(t)$ is determined by the two contributing terms describing the evolution of the density and the curvature, determined by Eqs.~\eqref{eq:EQdensitiesLCDMfull}.
A consequence of the equivalence principle of \acidx{GR} is that observers, freely falling in a gravitational potential, reside in a local inertial system, where \acidx{SR} applies, that is, they perceive flat space (see e.g., \citet{Weinberg1972,Weinberg2008}, \citet{Peacock1999} or \citet{Fliessbach2016}). Moreover, a consequence of this is, that space appearing flat to comoving \acidx{FLRW} observers is irrespective of the energy density of the model universe. Thus, space appears flat to comoving observers in open, closed and flat universes, justifying $\syidx{SYMed}_k=0$ and $\syidx{SYMdenspar}_{k,0}=0$ in the \glo{LCDMmodel} and provides the reason for the observation of the flatness of space, not necessarily connected to the \glo{criticaldensity}. This again suggests that the curvature term in the \glo{friedmannequation} Eqs.~\eqref{eq:EQfriedmannsgeneralNLK} and ~\eqref{eq:EQfriedmannLCDMnbg} does not express the global spatial curvature, as determined by the Einstein equations \eqref{eq:iCDMEQEinsteinNL}. 

Therefore, the interpretation of the curvature term in Eq.~\eqref{eq:EQfriedmannsgeneralNLK} is still a pending  question, as the term is not related to $\syidx{SYMed}_k$ and $\syidx{SYMdenspar}_{k,0}$, respectively, which express the observed flatness of space, by us as free-falling \acidx{FLRW} observers\footnote{The fact that we are not perfect \acidx{FLRW} observers has ramifications, which we study in Section~\ref{sec:resultsowCDM}.}.

\subsection{The Initial Conditions of FLRW Universes}\label{sec:FLRWgeometry}

Customarily, the geometry (open, closed or flat) of a model universe is explained based on the energy density of the background universe relative to the \glo{criticaldensity}. We present a more general definition based on the \acidx{IC} of the background universe, comprising of the initial densities in the early Universe, as well as the initial (post \glo{bigbang}) expansion rate.

The expansion rate for a universe at \glo{criticaldensity} is described by the \glo{friedmannequation} with vanishing curvature term
%
\begin{equation} \label{eq:EQfriedmannsgeneralFlat}%
	H^{2} =
	\frac{8 \pi G}{3 c^{2}} \rho_{\text{crit}},
\end{equation}
This relation between expansion rate $H$ and $\rho_{\text{crit}}$ holds true for the entire cosmic time. In particular, also in the very first moments after the big bang. More precisely, the initial boost in the expansion rate is supposed to be provided immediately after the big bang. By the time we can apply \acidx{GR}, we can define an initial expansion rate $H_{\text{ini}}$ (in the language of Lema\^{i}tre, we could call it here \q{the primeval expansion rate}). As of this cosmic point in time, when we can meaningfully talk about the Friedmann equation, the Universe experienced its first deceleration phase (see e.g. \citet{Harrison2000}), and the metric would appear flat for a comoving FLRW observer.

On the other hand, we can express the critical density for a flat universe by
%
\begin{equation} \label{eq:EQcritdensbb}%
	\rho_{\text{crit}} = \frac{3 H^2_{\text{ini}} c^{2}}{8 \pi G},
\end{equation}
which is simply Eq.~\eqref{eq:EQfriedmannsgeneralFlat} rearranged to express $\rho_{\text{crit}}$ at the considered initial point in time. This defines the \qc{critical expansion rate}{,}  for a given initial density as
%
\begin{equation} \label{eq:EQcritHbb}%
	H^2_{\text{crit}}=\frac{ 8 \pi G }{ 3 c^2 } \rho_{\text{ini}}.
\end{equation}
We can interpret this relation as follows. Given an arbitrary initial energy density $\rho_{\text{ini}}$ of a universe, originating from the big bang, the primeval expansion rate $H_{\text{ini}}$ has to be specifically fine-tuned, in order to fulfill criterion \eqref{eq:EQcritHbb} describing a universe with flat geometry, that is $H_{\text{ini}} = H_{\text{crit}}$. 

However, there are no comprehensible arguments, why the big bang should be restricted to this exclusive fine-tuned value for $H_{\text{ini}}$. In case $H_{\text{ini}}$ is less than $H_{\text{crit}}$ (or in other words, the $\rho_{\text{ini}}$ is higher than $\rho_{\text{crit}}$, in order to fulfill Eq.~\eqref{eq:EQcritHbb}), the evolution of the universe is described by a closed geometry. An open geometry is determined by $H_{\text{ini}}$ greater than $H_{\text{crit}}$ (or $\rho_{\text{ini}}$ being lower than $\rho_{\text{crit}}$, fulfilling Eq.~\eqref{eq:EQcritHbb}).

Limiting ourselves to a flat geometry and given the energy densities as deduced by the measurements of the \acidx{CMB} (e.g., by \citet{Collaboration2020}) in \acidx{LCDM}, it thereby ignores from the outset a broad range of possible initial expansion rates $\syidx{SYMH}_{\text{ini}}$. Hence, the assumption of a flat geometry in the \glo{flrwmetric} does not cover those initial expansion rates $\syidx{SYMH}_{\text{ini}}$, which would lead to subcritical and supercritical universes, where, as shown in Sec.~\ref{sec:curvatureFLRW}, freely falling, comoving observers likewise perceive flat space. 

\section{A \boldmath$\Lambda$CDM Extension to incorporate the Initial Conditions}\label{sec:EnhancedFriedmann}

In Sec.~\ref{sec:curvatureFLRW}, we argue that a prospective observation of flat space does not necessarily imply a universe at \glo{criticaldensity}, since the curvature terms $\syidx{SYMed}_{k}$ and $\syidx{SYMdenspar}_{k}$ in  Eqs.~\eqref{eq:EQfriedmannLCDMnbg} and \eqref{eq:EQclosureLCDMfull}, respectively, vanish in the local inertial frame of comoving \acidx{FLRW} observers, giving the reasons for the flatness of space as measured by the observations of the \acidx{CMB}. 
In order to interpret the curvature term in Eq.~\eqref{eq:EQfriedmannsgeneralNLK}, we associate the general concept of \acidx{DE} with the geometry of the \glo{flrwmetric}~\eqref{eq:iCDMEQflrwbgk} and use the subscript \q{de} for the quantities describing the dynamics of expansion in the forthcoming.

In Sec.~\ref{sec:FLRWgeometry} we argue that the \acidx{IC} of the background universe are given by the initial expansion rate $H_{\text{ini}}$ and the initial densities. Based on the \acidx{CMB} measurements, within the \acidx{LCDM} model, the initial densities are determined to high precision (see also \citet{Foidl2024}). For this reason, it is sufficient to incorporate $H_{\text{ini}}$ into the \acidx{LCDM} formalism. 
Let us proceed in our approach by considering the following parameters
%
\begin{equation} \label{eq:EQclosureLCDMiCDM}%
	1 = \Omega_{r,0} + \Omega_{b,0} + \Omega_{\text{CDM},0}  + \Omega_{k,0} + \Omega_{\aCC,0},
\end{equation}
where the operationally defined density parameter of the geometrical curvature $\Omega_{\gc}$ takes the place of $\Omega_{\Lambda}$. For the sake of completeness, we include $\Omega_{k} = 0$ in the equation, in order to express the perceived flatness of space. In the same way as in \acidx{LCDM}, $\Omega_{k,0}=0$ describes the \qemph{flatness of space}, based on novel arguments, though, as it appears to us in our local reference frame as comoving \acidx{FLRW} observers.

In order to proceed with our approach, in an inflationary big bang cosmology, we allow for the following simplification.
We analyze the evolution of cosmological models by the time inflation has ended and we call the expansion rate at the end of inflation \q{primordial expansion rate} (in analogy to the primordial power spectrum in structure formation). Detailed information of the exact evolution of $H$ prior to this point is not required.

We recognize from Eq.~\eqref{eq:EQclosureLCDMiCDM} 
%
\begin{equation} \label{eq:EQOmegaH}%
	\Omega_{\gc,0} = 1 - \Omega_{\rphys,0},
\end{equation}
where $\Omega_{\rphys,0}$ denotes the sum total of the density parameters ($\Omega_{\text{CDM},0}$, $\Omega_{b,0}$, $\Omega_{r,0}$) of all physical contributions to the energy budget of the universe (CDM, baryons, radiation; without consideration of the operationally defined contributions, $\syidx{SYMdenspar}_{k,0}$ and $\syidx{SYMdenspar}_{\aCC,0}$, to the energy momentum tensor $T_{\mu \nu}$). 

We rewrite the \glo{friedmannequation} \eqref{eq:EQfriedmannsgeneralNLK}, by multiplying Eq.~\eqref{eq:EQfriedmannsgeneralNLK} by $\syidx{SYMa}^2(t)$ and divide it by 2, and where we use the total amount of energy $(4\pi/3) \syidx{SYMed}_{\rphys} \syidx{SYMa}^3$ in the sphere of radius (scale factor) $\syidx{SYMa}$, in units of the \glo{criticaldensity} (see Eq.~\eqref{eq:EQOmegaH}), instead of the respective energy density $\syidx{SYMed}_{\rphys}$, which yields
%
\begin{equation} \label{eq:EQfriedmanngeneralNLintegral}%
	\frac{1}{2}\dot{a}^{2}(t) - \frac{ G \Omega_{\rphys}}{a} = \kappa,
\end{equation}
with the constant $\syidx{SYMcurvpar}$.

We use the \acidx{EOS} parameter of $\syidx{SYMed}_{\gc}$ to parameterize the dynamics of the expansion, determined by the curvature index in the \glo{flrwmetric}, as a function of the sum total of the energy densities of the cosmic components in the Universe $\syidx{SYMdenspar}_{\rphys}$, which retains the customary \acidx{LCDM} formalism, while we are only left to adapt the computation of $\syidx{SYMeos}_{\gc}$. 

So, we use now Eq.~\eqref{eq:EQfriedmanngeneralNLintegral} to determine $w_{\gc}$ as follows. 
Since $\kappa$ is a constant, the evolution of the first term is determined by the evolution of the second term, given by
%
\begin{equation} \label{eq:EQevolEpot}%
	\frac{ d }{ da } \left( \frac{ G \Omega_{\rphys}  }{ a } \right) \propto - \frac{ \Omega_{\rphys} }{ a^2 }.
\end{equation}
which is, unsurprisingly, equivalent to Eq.~\eqref{subeqn:EQdensitiesLCDMfk}. (In what follows, we do not need detailed prefactors.)
We use relation \eqref{eq:EQevolEpot} in Eq.~\eqref{eq:EQdensevolgen} which yields
%
\begin{equation} \label{eq:EQevolrhopot}%
	\rho_{\gc} \propto a^{-2\Omega_{\rphys,0}},
\end{equation}
and equate the exponents in \eqref{eq:EQevolrhopot} and $\rho_{k} \propto a^{-3(1+w)}$ [see Eq.~\eqref{eq:EQdensevolgen}], rearranged to express $w_{\gc}$ it reads as
%
\begin{equation} \label{eq:EQfriedmannparamw}%
	w_{\gc\text{\sprim}} = \frac{2}{3} \Omega_{\rphys,0} - 1,
\end{equation}
where $w_{\gc\text{\sprim}}$ is constant.

The significant property of \eqref{eq:EQfriedmannparamw} is the fact that, in general, it does not yield the \acidx{EOS} of a cosmological constant. Only for an empty universe, we do \qemph{exactly} get $w_{\gc\text{\sprim}}=-1$. In fact, this is in nice accordance with the original \textit{empty} de Sitter universe solution (\citet{Sitter1917}) and the expectations from a kinematically determined universe: If the universe is empty, there is no global curvature of space, and hence no deceleration, such that $H=const$, resulting in an exponential growth of the scale factor.

As soon as we have cosmic components, $\Omega_{\rphys,0} > 0$,  the \acidx{EOS} parameter in \eqref{eq:EQfriedmannparamw} fulfills $w_{\gc\text{\sprim}} > -1$.
If we choose the same matter content as the $\Lambda$CDM concordance model, i.e., using $\Omega_{\rphys,0} \approx \Omega_{m,0} \simeq 0.3$, Eq.~\eqref{eq:EQfriedmannparamw} yields $w_{\gc\text{\sprim}} \simeq -0.8$, thus there is a weak deceleration, compared to a model with $w_{\Lambda} = -1$. Still, the two \acidx{EOS} parameters are close numerically and in terms of their phenomenological impact onto the expansion history (see Section~\ref{sec:resultsiCDM}).

On the other hand, if we assume a universe at \glo{criticaldensity}, that is $\syidx{SYMdenspar}_{\rphys,0} =1$, yields $\syidx{SYMeos}_{\gc\text{\sprim}}=-1/3$, which is the \acidx{EOS} parameter of the spatial curvature used in the \acidx{FLRW} formalism, and which is restricted to the fine-tuned case of a universe at \glo{criticaldensity}. This is exactly what we expect.

To retain \acidx{LCDM}'s formalism and Eq.~\eqref{eq:EQdensevolgen}, we make a distinction of cases. We apply the constant value of $w_{\gc\text{\sprim}}=-1/3$ to flat and closed geometries, as is also the case in $\Lambda$CDM (with $w_k=-1/3$). For open geometries, we apply Eq.~\eqref{eq:EQfriedmannparamw}.
In summary, we have
%
\begin{subequations} \label{eq:EQfriedmannparamsSingle}%
	\begin{empheq}{align}
	w_{\aCC,\sini} &= -\frac{1}{3} - \Theta(\syidx{SYMdenspar}_{\aCC,0}) \frac{2}{3} \syidx{SYMdenspar}_{\aCC,0} \label{subeqn:EQfriedmannparamsSingle1}\\
\nonumber\\
\syidx{SYMdenspar}_{\aCC,0} &= 1 - \syidx{SYMdenspar}_{\rphys,0}\label{subeqn:EQfriedmannparamsSingle2}
	\end{empheq}
\end{subequations}
where in Eq.~\eqref{subeqn:EQfriedmannparamsSingle1} $\Theta$ is the Heaviside function.
The \acidx{EOS} parameter $w_{\aCC,\sini}$ is a function of $\Omega_{\aCC,0}$ [see Eq.~\eqref{subeqn:EQfriedmannparamsSingle2}], and therefore a \qemph{constant} for a given total of energy densities. We use the additional subscript \q{\sini} to express that this is the initial value for the \acidx{EOS} parameter in the early Universe. In Sec.~\ref{sec:largescale}, we will show that in the regime of nonlinear structure formation, the EoS will morph from a constant value to a time-dependent (or scale-factor-dependent) function. The Heaviside function separates the two regimes of super- and subcritical model universes. The first term corresponds to deceleration due to the critical density. The factor after the Heaviside function applies to subcritical universes only, where $w_{\aCC,\sini}$ falls below $-1/3$, that is a decreasing \acidx{EOS} parameter implies less deceleration, as it should.
In addition, we retain $\syidx{SYMdenspar}_k = 0$ to express the perceived flatness of space, in our local inertial system (as comoving \acidx{FLRW} observers), just the same way as in \acidx{LCDM}. 
This is essential to linear perturbation theory applied in \acidx{LCDM}, when, for example, calculating the \acidx{CMB} temperature spectrum, as the notion \q{curvature} herein refers to spatial curvature in the Einstein equations (see e.g., \citet{Ma1995,Weinberg2008,Coles2002,Mukhanov2005,Dodelson2003,Peebles1993}), and not to the geometry of the \glo{flrwmetric}. A description of how the spatial curvature affects the \acidx{CMB} temperature spectrum can be found in many textbooks covering structure formation, see the aforementioned references. But there is a degeneracy for the impact of spatial curvature on the \acidx{CMB} spectrum with the \glo{cosmoconstant} \CCLambda{} or \glo{darkenergy}, respectively; see for example, \citet{Hu2002}. We will get back to this point in Sec.~\ref{sec:resultsowCDM}. 

Finally, the Friedmann equation reads
%
\begin{subequations} \label{eq:EQfriedmannenhanced}%
	\begin{empheq}{align}
		H^2(t) &= \frac{8 \pi G}{3 c^{2}} \left[ \rho_{r}(t) + \rho_{b}(t) + \rho_{\text{CDM}}(t) + \rho_{\aCC}(t)  
		\right] \label{subeqn:EQfriedmannenhanced1}\\
		\rho_{\aCC}(a) &= \Omega_{\aCC,0} \rho_{\text{crit},0} \: a^{-3(1+w_{\aCC,\sini})} \label{subeqn:EQfriedmannenhanced2},
	\end{empheq}
\end{subequations}
where Eq.~\eqref{subeqn:EQfriedmannenhanced2} now describes the evolution of $\rho_{\aCC}$ as a function of scale factor $a$ for a constant $w_{\aCC,\sini}$, given by Eqs.~\eqref{eq:EQfriedmannparamsSingle}. The present-day critical density $\syidx{SYMed}_{\text{crit},0}$ is defined in \eqref{eq:iCDMEQcritdenst}. 

\begin{figure} [!htb]%
	\includegraphics[width=1.0\columnwidth]{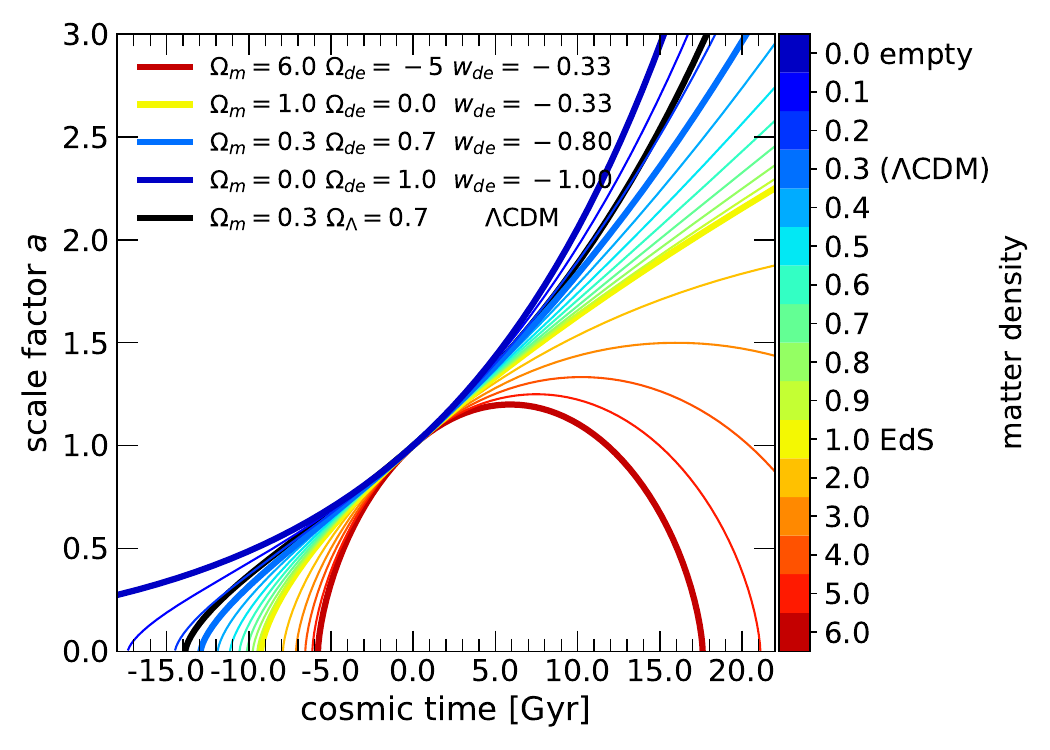}
	\caption[Expansion histories of model universes within the $\Lambda$CDM extension]%
	{\tb{Expansion histories of model universes within the \boldmath$\Lambda$CDM extension.} The color-coded curves display the expansion history of individual model universes applying Eq.~\eqref{eq:EQfriedmannenhanced}, for models with supercritical density (dark red), to the \acidx{EdS} model (yellow), to the empty de Sitter universe (dark blue). The black curve indicates the expansion history of the original $\Lambda$CDM model with $\Omega_{m} = 0.3, \Omega_{\Lambda}=0.7$, assuming a cosmological constant, i.e. $w_{\Lambda} = -1$. Comparing \acidx{LCDM} to the model shown for the same matter density and $\Omega_{\aCC} = 0.7$ (thick light-blue curve), we see a great similarity. This model displays the same characteristic s-shape, indicating the transition from decelerated to accelerated expansion, because of $w_{\aCC} = -0.8$ being \q{close} to a cosmological constant, given the relatively low energy density observed in the Universe.
	}%
	\label{fig:Friedmann-wih-closed}%
\end{figure}%
Figure \ref{fig:Friedmann-wih-closed} finally displays the evolution of the scale factors of model universes with various matter densities, color-coded by density parameter $\Omega_m$, whose value refers to the present, within our \newmodel{}, from reddish for universes with closed geometry; the yellow one has exactly the critical density, that is the \acidx{EdS} universe, which separates the supercritical from the subcritical universes, which go from light green to deep blue for the empty universe.
The curves between the yellow and the dark blue curve depict the evolution of subcritical models with matter densities between the \acidx{EdS} model with critical density (solid yellow curve), and the empty model (solid dark blue curve). We can see that the cosmological models transition uniformly between these two limiting model cases, corresponding to decreasing mass density, and \q{approaching} the exponential curve of the empty model, exactly as expected for kinematically determined universes in \acidx{GR}. In fact, a universe filled with a cosmological constant to critical density displays the same evolution as the expectations for an empty universe, in the former by pressure balancing gravity\footnote{This scenario also applies to the inflationary phase of the Universe, when $p \approx -\rho$ of the inflaton field  dominated the Universe, resulting in an exponential growth of the scale factor.}, in the latter by vanishing gravity in an empty universe.\par
The black solid curve indicates the evolution of the \glo{LCDMmodel} with $\syidx{SYMdenspar}_m = 0.3$ and the \glo{cosmoconstant} $\syidx{SYMdenspar}_{\Lambda} = 0.7$. Comparing it to the solution obtained by our \newmodel{} with $\syidx{SYMdenspar}_{\aCC} = 0.7$ and $\syidx{SYMeos}_{\aCC} = -0.8$ for a model universe with identical amount of matter $\syidx{SYMdenspar}_m = 0.3$ (thick solid blue line), we recognize the similarity between both models. Both display the typical s-shape in the evolution of the scale factor, indicating the transition from decelerated to accelerated expansion, known from \acidx{LCDM}. There is a difference in the age between the two world models, however. We discuss this finding and other features, along with the results of exact calculations in Sec.~\ref{sec:resultsiCDM}.
Moreover, the evolution of supercritical model universes with matter densities above the critical density match the expectation, as well. They display no deviations from conventional computations, with respect to the evolution of their scale factors.

This approach shows a significant difference to the \glo{cosmoconstant} \CCLambda{}, that is considered a physical content of the Universe. But unlike the \glo{cosmoconstant} \CCLambda{}, $\syidx{SYMed}_{\aCC}$ now \qemph{does not} contribute to the energy budget of the Universe, instead it refers to a kinematic effect, described by an effective \acidx{DE} component emerging from the geometry of the \glo{flrwmetric}. As such it is considered an operational contribution to the energy momentum tensor [Eq.~\eqref{eq:EQfriedmannLCDMnbg}] and is not regarded as a physical contribution to the Universe. Just in the same way as $\syidx{SYMed}_k$ in \acidx{LCDM}, obviously. Though it seems to be only a minor modification, it is a significant difference to the \glo{LCDMmodel}, as the Universe, in this approach, is considered an open universe with subcritical energy density, as the sum total of the energy densities of the physical contributions (radiation and matter) to the energy budget of the universe is below \glo{criticaldensity}.
Finally, $\syidx{SYMdenspar}_{k,0} = 0$ refers to the perceived flatness of space in the local inertial system of the comoving \acidx{FLRW} observer, in contrast to \acidx{LCDM}, where it is interpreted as the global flatness of space.

\section{The Expansion History in the Nonlinear Regime}\label{sec:largescale}
%
In Sec.~\ref{sec:EnhancedFriedmann}, we presented an extension to \acidx{LCDM}, which abandons the cosmological constant $\Lambda$ in favor of an approach that describes the DE phenomenology as a kinematic effect, induced by the primeval expansion rate $H_{\text{ini}}$ post big bang in relation to the initial density, in order to extend the \glo{LCDMmodel} beyond the fine-tuned case of a Universe at \glo{criticaldensity}.

In $\Lambda$CDM, the nonlinear stage of structure formation is assumed to have no impact on the evolution of the background universe. In this section, we reassess the possibility of such an impact.
More precisely, we investigate the impact of the formation of the cosmic web, in particular of its voids, onto the expansion history. We can show that there is an impact which is minor, but significant enough to explain the Hubble tension problem, on which we elaborate in detail below. 

In order to motivate our approach, we  
reconsider first the ideas around the issue of a backreaction from structure formation, particularly in the late stages of the evolution of the cosmic web, since by then voids dominate the volume of the Universe.\par

\subsection{The evolution of the voids}\label{sec:evolvoids}
%
The pioneering work of \citet{Icke1984} showed that voids evolve into spherical shape and become distributed homogeneously during the formation of the cosmic web. The following works by \citet{Icke1987}, \citet{Weygaert1989} and \citet{Weygaert1994} presented analytical descriptions of the evolution of the cosmic web, based on the Voronoi tesselation (\citet{Voronoi1908}, see also \citet{Okabe2000}).  \citet{Icke1991} confirmed the correctness of this approach by comparing it to observations. Analytical approximations are presented in \citet{Icke2001}, which describe the evolution of the individual components of the cosmic web as Voronoi features, which we display here
%
\begin{subequations} \label{eq:EQevolweb}%
	\begin{empheq}{align}
		m_v =& e^{-3\theta}\\
		m_w =& 3e^{-2\theta}\left(1 - e^{-\theta}\right)\\
		m_f =& 3e^{-\theta}\left(1 - e^{-\theta}\right)^2\\
		m_n =& \left(1 - e^{-\theta}\right)^3,
	\end{empheq}
\end{subequations}
where $\theta$ is a measure of time, with $\theta \propto t^{2/3}$ in an \acidx{EdS} universe. The quantities $m_x$ denote the mass fractions of the individual components of the cosmic web: (a) $m_v$ for the voids, (b) $m_w$ for the walls, (c) $m_f$ for the filaments and (d) $m_n$ for the nodes.\par
As voids became a subject of interest, \citet{Colberg2005} and \citet{Shandarin2006} derived density profiles, shapes and sizes from cosmological simulations. \citet{Ricciardelli2013} showed that voids display a universal density profile. \citet{Cautun2014} derived the evolution history of the individual components of the cosmic web from the Millennium simulation of \citet{Springel2005a}.\par
%
\subsection{The backreaction problem}\label{sec:averagingproblem}
%
The expansion history of the Universe in the \glo{LCDMmodel} is determined by the solution of the Friedmann equations, applying the average energy density of the background universe. This is a valid approach, as long as the distribution of the energy densities can be considered homogeneous and isotropic, that is in the early Universe. However, the cosmological principle is not valid anymore in the late Universe on spatial scales of the order of approximately Gpc and below. Thus, averaging the density of the Universe and solving the Friedmann equations by neglecting this nonlinear structure formation seems to be problematic. This is known in cosmology as the \qemph{averaging problem}. The theoretical background of this problem arises from the nonlinearity of the Einstein equations
%
\begin{equation} \label{eq:EQEinsteinE}%
	E_{\mu\nu} = \frac{8 \pi G}{c^4} \: T_{\mu\nu},
\end{equation}
first pointed out by \citet{Ellis1983}.
Averaging the metric in an inhomogeneous environment first, before solving the Einstein equations, will lead to an additional correction term $C_{\mu\nu}$ in the averaged Einstein equations:
%
\begin{equation} \label{eq:EQEinsteinAVG}%
	\langle E_{\mu\nu} \rangle = \frac{8 \pi G}{c^4} \: \langle T_{\mu\nu} \rangle + C_{\mu\nu}.
\end{equation}
The questions to be addressed are the following: (a) can this term reach relevant orders of magnitude to influence the expansion history of the Universe? (b) does perturbation theory eventually break down?\par
In $\Lambda$CDM it is assumed that the correction term $C_{\mu\nu}$ is negligible. On the other hand, if this term would be found to impact the expansion history, it would be seen as a \qemph{backreaction process}, caused by inhomogeneities onto the background universe.\par
The questions (a) and (b) have been explored in the literature, providing basically the following answers: in case the size of perturbations is much less than the (time-dependent) Hubble sphere, the correction term does not have substantial impact on the expansion history and can be neglected. This means it cannot explain the phenomenology of the accelerated expansion, see e.g. \citet{Kolb2005a}, \citet{Notari2006}, \citet{Li2007}, \citet{Raesaenen2006a}, \citet{Buchert2000,Buchert2001}, 
\citet{Kwan2009}, \citet{Paranjape2009}.

\citet{Paranjape2009} explored the averaging problem and its importance in cosmology in a mathematically rigorous way. 
They constructed a toy universe based on the Lema\^{i}tre-Tolman-Bondi (LTB) metric (\citet{Lemaitre1933,Lemaitre1997,Tolman1934,Bondi1947}). The key property of this metric is that it allows radial inhomogeneities, while preserving isotropy. Their model was centered at an over-dense spherical region -- a model for a formed dark matter halo --, surrounded by an under-dense shell. The remaining part of the universe was regarded homogeneous out to the Hubble sphere. They found that the value of the correction term $C_{\mu \nu}$ induced by the inhomogeneity is too small, in order to have an impact onto the expansion history, justifying the assumptions of $\Lambda$CDM. However, while the toy model analyzed in \citet{Paranjape2009} perfectly fits to the regime of halo formation and virialization, it does not apply to the late stages of cosmic web formation, once voids dominate the volume of the Universe.\par
\citet{Buchert2011} also investigated the averaging problem, based on the LTB metric, and finds that the global backreaction in a flat LTB model vanishes in a spherical domain. This result is compatible with the interpretation of the Universe to appear homogeneous, if we chose a large enough spatial scale, i.e. $\gtrsim 100$~Mpc at the present (\citet{Hogg2005,Scrimgeour2012}), justifying the assumptions in $\Lambda$CDM. Moreover, \citet{Buchert1997} have shown that in 3D-torus architecture global backreaction vanishes, as well, meaning that in cosmological simulations with periodic boundary conditions backreactions cannot appear.
%
\subsection{The backreaction from voids}\label{sec:avgproblem}
%
Although backreaction from the averaging problem has been shown to vanish in LTB models, \citet{Buchert2015} demonstrated that there is no proof that backreaction by inhomogeneities is negliglible, in general, in cosmology. Furthermore, it has been pointed out e.g. by \citet{Amendola2015,Kolb2005a} that the result of the averaging problem is very sensitive to the way the average is performed.\par
In this paper, we reconsider the averaging problem as applied to the late stages, when voids dominate the volume of the Universe. 
More precisely, we distinguish the impact of density vs volume. In the late stages of nonlinear structure formation, the Universe is dominated in volume by a homogeneous distribution of spherical voids of very low density, embedded in a highly over-dense region, which occupies only a minor fraction of the volume of the Universe. This picture is compatible with the general-relativistic \q{separate universe conjecture} (\citet{Weinberg2008}), which states that a spherically symmetric region in a homogeneous and isotropic universe behaves like a mini universe. 
There are three approaches, which each can support the conjecture that voids as the dominating volume fraction of the Universe influence the expansion history: a) a phenomenological approach; b) a backreaction process; c) results of simulations; supported by d) results from observations, as discussed next.\par
\noindent\tb{a) phenomenological approach}\par
%
Following the separate universe conjecture, every void can be considered a mini universe, having its own expansion rate. Moreover, we do not consider the evolution of a single void, but in a holistic view, we consider the entire Hubble sphere.\par
The expansion rate of the Universe decreases with time, due to the action of gravity. This deceleration is determined by the gravitational forces $F_G$ between any two points ($p_1,p_2$) in the homogeneous universe. Following Birkhoff's theorem (\citet{Birkhoff1923,Jebsen1921,Jebsen2005,Maciel2018}), gravitational forces in a homogeneous sphere are proportional to the distance $r$ between any point $p$ from the center $p_c$: $F_G \propto r(p,p_c)$. 
Considering points $p_x$ at the edge of a over-dense region surrounding a very under-dense void, yields the usual Newtonian law of gravity $F_G \propto 1/r^2(p_x,p_c)$. This can be applied to the scenario, where voids are embedded in an over-dense environment, showing that deceleration in the voids is less pronounced than in the surrounding cosmic web. Alternatively, this can also be seen by the Birkhoff's theorem for the spherical low-density voids, yielding a lower gravitational force within the voids compared to their higher-density homogeneous environment. This means that voids are experiencing an accelerated expansion, compared to their higher-density homogeneous environment. As long as voids do not dominate the volume of the Universe, this expansion can be compensated by the \q{congestion} of the over-dense walls and filaments. As soon as voids begin to dominate the volume of the Universe, this compensation fails. Voids dominate the volume of the Universe and therefore the expansion of the Universe, as it is directly connected to the volume of the voids.\par
\noindent\tb{b) backreaction process}\par
%
The solutions to the averaging problem which we present in Sec.~\ref{sec:averagingproblem} consider the size of inhomogeneities clearly smaller than the Hubble radius. Therefore, it is always possible to find a minimum spatial scale, above which the Universe can be considered homogeneous, yielding no impact on the expansion history.\par
However, as studies have shown, in case perturbations are of the size of the Hubble radius or even larger (\q{superhorizon modes}), the correction term can grow to significant values, and therefore may impact the expansion history of the Universe, which appears as a classical (zero-momentum) background (see e.g. \citet{Carloni2008},  \citet{Martineau2005}, \citet{Barausse2005}, \citet{Kolb2005a},  \citet{Parry2006}, \citet{Kumar2008}, \citet{Hirata2005}, \citet{Flanagan2005}, \citet{Raesaenen2006}, \citet{Geshnizjani2005}). This topic is subject to controversial discussions, and it has been even questioned if such huge perturbations exist at all (see e.g.  \citet{Carloni2008}, \citet{Stoeger2007}, \citet{Wetterich2003},  \citet{Calzetta2001}, \citet{Siegel2005}, \citet{Gasperini2009},  \citet{Gruzinov2006}, \citet{Notari2006}, \citet{VanAcoleyen2008}).\par
Yet, relativistic perturbation theory does predict superhorizon perturbations, and as the Hubble sphere grows over time, more and more such perturbations enter that sphere (perturbations leave the sphere, once the scale factor grows exponentially and $H=const$). Either way, it appears that these superhorizon perturbations are far too subdominant to source a backreaction process onto the background evolution, in order to explain the phenomenology of the accelerated expansion.\par
In our model, we take into account the evolutionary stage of the cosmic web, once voids dominate the volume of the Universe, that is low-density spherically symmetric voids are embedded in a highly over-dense homogeneous background. 
Furthermore, we consider the impact of the entire \glo{cosmicweb} on the backreaction process, meaning we consider the entire Hubble sphere. Therefore, we can regard the entirety of all voids as \qemph{one} huge extreme low-density region -- appearing as a classical (zero-momentum) background, see for example, \citet{Barausse2005}, \citet{Kolb2005a}).\par
\noindent\tb{c) backreaction via spatial averaging}\par
%
Let us add another aspect of the backreaction.
In \citet{Buchert2015} it was shown that, in general, there is no proof that inhomogeneities are negligible in background cosmology. Nevertheless, the success of the \glo{LCDMmodel} suggests that any impact of inhomogeneities on the expansion rate should be weak. However, in \citet{Wiltshire2007,Wiltshire2011} it was shown that particularly in the late stages of the evolution of the cosmic web, averaging and coarse-graining is a problematic procedure.\par
\citet{Racz2017} performed a cosmological N-body simulation, integrating the Newtonian equations, as is customary in cosmological simulations (e.g., \citet{Springel2005a}), with a changing general-relativistic metric, which is calculated from spatially averaged quantities. The result indicated that for a typical spatial scale of the averaging procedure, an impact onto the expansion history with deviations from the results based on the FLRW metric is possible.

Before we close this section, we reiterate that all studies indicate that the backreaction onto the background evolution, including the backreaction from voids, cannot \q{explain away} the phenomenology of accelerated expansion.
However, by including the minor backreaction of voids that we find, we can possibly resolve the Hubble tension problem, as shown below.

\noindent\tb{d) results from observations}\par
%
As discussed in a later section, recent work by e.g., \citet{Dainotti2022,Dainotti2021,Dainotti2022a,Bargiacchi2023,Bargiacchi2023a,OColgain2021} and \citet{Krishnan2021a}, indicate that the evolution of the expansion rate $H$ at low redshift is in conflict with the assumption of a cosmological constant. These results may indicate that the origin of the Hubble tension is likely to be in cosmology and not in local measurement issues, where a dynamical DE component, our proposed model, or a backreaction from the late stages of structure formation might each a priori constitute a solution.

\subsection{Including Backreaction into the \boldmath$\Lambda$CDM Extension}\label{sec:finalequations}

There is a straightforward approach to incorporate the backreaction from the cosmic web and its voids into the \newmodel{}, and at the same time to quantify the corresponding impact onto the expansion history, as follows. We apply the hydrodynamical and geometrical models from \citet{Icke1984}, \citet{Icke1987} and \citet{Icke2001} for the formation of the cosmic web and extract the relevant quantities from the analysis by \citet{Cautun2014} of the Millennium simulation. The advantage of this procedure is the fact that the large-scale structure of the Universe in our model is informed by cosmological $\Lambda$CDM simulations, which are in good agreement with observations, that is with large galaxy surveys on scales of the cosmic web.\par
In the \newmodel{}, which we presented in Sec.~\ref{sec:EnhancedFriedmann}, the cosmological constant $\Lambda$ is replaced by an \q{operational dark energy}, where the associated EoS is given in Eqs.~\eqref{eq:EQfriedmannparamsSingle}. As previously mentioned, the expansion of the voids is directly connected to the expansion of the Universe by their dominance in volume and their low density.
Therefore, it is straightforward to express the impact from voids onto the expansion history through the EoS of $\rho_{\aCC}$. As a result, the initially constant EoS parameter $w_{\aCC,\sini}$ morphs into a time-dependent function. This makes sense, because voids become important only at the later stages of structure formation.
To quantify this effect, we used the results derived by \citet{Cautun2014}, who analyze the evolution of the cosmic web and who derive the evolution of the mass fractions and volume fractions for the individual components of the cosmic web (i.e., voids, filaments, walls, nodes), using the Millennium simulation by \citet{Springel2005a}. \citet{Cautun2014} find that voids dominate the volume of the Universe as of a redshift $z \sim 5$. The analytical model \eqref{eq:EQevolweb} from \citet{Icke2001} shows that the matter fraction of the voids evolves linearly in redshift, to a good approximation, in the late stages of the evolution of the cosmic web. We applied both findings to our Eqs.~\eqref{eq:EQfriedmannparamsSingle}, in order to derive the following linear approximation for the evolution of the \acidx{EOS} parameter of $\rho_{\aCC}$ as a function of the scale factor as follows. We use Eqs.~\eqref{eq:EQfriedmannparamsSingle} to obtain the 	$\syidx{SYMeos}_{\aCC,\sini}$ (from parameters reported by the \citet{Collaboration2020}) and $\syidx{SYMeos}_{\aCC,0}$ (from the results reported by \citet{Cautun2014}). In the linear evolution of redshift from $z=5$ to $z=0$, we use $z+1=1/\syidx{SYMa}$ to express $\syidx{SYMeos}_{\aCC}$ as a function of scale factor $\syidx{SYMa}$, which yields
%
\begin{equation} \label{eq:EQwbackreactionCautun}%
\begin{aligned}%
	w_{\aCC}(a) &= \begin{cases}%
		w_{\aCC,\sini} \text{~~~when~} a \leq 1/6\\%
		w_{\aCC,\sini} + \left[\frac{(1/a-1) - 5}{5}\right] (w_{\aCC,\sini} + 0.9)\end{cases}\\%
		~&\text{~~~~~~~~~~~~~~~~~~~~~~when~}a > 1/6.
\end{aligned}
\end{equation}
Thus, the \acidx{EOS} parameter $w_{\aCC}$ becomes a function of cosmic time, or scale factor, respectively, as indicated in the formula, once the scale factor has reached the threshold determined by the aforementioned analysis.
Using the density parameters of our model with $\Omega_{\aCC,0} = 0.7$ and $\Omega_{m,0} = 0.3$, we get for the present-day EoS parameter $w_{\aCC}(1) \sim -0.9$. Comparing this result with the constant \acidx{EOS} parameter $w_{\aCC,\sini} = -0.8$ of Sec.~\ref{sec:EnhancedFriedmann}, we see that the impact of voids leads to a more negative \acidx{EOS} parameter, with a value closer to that of a cosmological constant. In fact, this is an expected outcome, given the dynamics of voids.

In order to reflect the correct evolution of the background, we stress that the time-dependence of the \acidx{EOS} parameter requires a forward-in-time-integration of the energy conservation equation \eqref{eq:EQeconsnbg} to determine the evolution of the energy density $\rho_{\aCC}$. Thus, Eqs.~\eqref{eq:EQdensevolgen} and \eqref{subeqn:EQfriedmannenhanced2}, respectively, cannot be used naively.
Regarding this important point, we refer the reader to \citet{Foidl2024}, where we discuss in detail computational characteristics for cosmological models with a dynamical dark energy component and the implications of a decreasing vs increasing \acidx{EOS} parameter.

We mentioned already that the impact of voids makes the \acidx{EOS} parameter $w_{\aCC}$ more negative, than without this impact. In fact, as the density within voids decreases while their volume increases, $w_{\aCC}(a)$ will converge asymptotically to the \acidx{EOS} parameter $w_{\aCC} = -1$ of a cosmological constant.

\section{Results for the wCDM Extension to \boldmath$\Lambda$CDM}\label{sec:resultsiCDM}
%
In order to calculate accurately the background evolution of our \newmodel{}, and to compare it carefully with the concordance $\Lambda$CDM model and with observations, we use the open-source cosmological Boltzmann code \acidx{CLASS} \footnote{The code CLASS is publicly available at \href{https://lesgourg.github.io/class_public/class.html}{https://lesgourg.github.io/class\_public/class.html}}, which is designed, not only to provide a user-friendly way to perform cosmological computations, but also to provide a flexible coding environment for implementing customized cosmological models. The modular concept makes it possible to enhance the code, without the risk of compromising existing functionality. The underlying concepts, including an overview of coding conventions, can be found in \citet{Lesgourgues2011}. 
This version uses the Planck 2018 cosmological parameters from \citet{Collaboration2020}, as the default parameter set. Additionally, the configuration provides different sets of precision configuration files, to reflect varying requirements on precision, which may be needed in the results or available computation time. The precision configuration with the highest accuracy is proofed to be in conformance with the Planck results within a level of $0.01\%$.\par
We use the fiducial parameters of the \glo{LCDMmodel} from \citet{Collaboration2020}, shown in Table \ref{tab:planck2018iCDM}, in our calculations. 
We properly implemented the equations of our \newmodel{} \eqref{eq:EQfriedmannparamsSingle},  \eqref{eq:EQfriedmannenhanced} and  \eqref{eq:EQwbackreactionCautun} into the background module of \acidx{CLASS}, as well as into the perturbation module, because we also calculated and compared the \acidx{CMB} temperature spectra between our \newmodel{} and concordance $\Lambda$CDM, as presented in the course of the next section.
{\belowcaptionskip=0pt 
	\begin{table}[!htbp] 
		\caption[Cosmological parameters used in the computations]{Cosmological parameters used in the computations}\label{tab:planck2018iCDM}
		\centering
		\begin{tabular}{lrl}
			\hline
			\hline
			\textbf{Parameter}        & \textbf{Value}            & \textbf{Comment} \\
			\hline
			H$_{0}$          & $67.556$         & \\
			T$_{\text{CMB}}$~[K]    & $2.7255$         & \\
			N$_{\text{ur}}$         & $3.046$          & \\
			$\Omega_{\gamma,0}$ & $5.41867 \times$10$^{-5}$  & derived from T$_{\text{CMB}}$\\
			$\Omega_{\nu,0}$    & $3.74847 \times$10$^{-5}$  & derived from N$_{\text{ur}}$\\
			$\Omega_{b,0}$      & $0.0482754$      & \\
			$\Omega_{\text{CDM},0}$   & $0.263771$       & \\
			$\Omega_{\aCC,0}$, $\Omega_{\Lambda,0}$& $0.687762$       & 
            wCDM, $\Lambda$CDM\\
            $\Omega_{k,0}$      & $0$&\\       
			$\tau_\text{{reio}}$     & $0.0925$         & \\			
			A$_{s}$          & $2.3 \times$10$^{-9}$       & \\
			n$_{s}$          & $0.9619$            & adiabatic ICs\\
			\hline
			\hline
		\end{tabular}
	\end{table}
}

In order to compare the results of our \newmodel{} to the concordance $\Lambda$CDM model, two \acidx{CLASS} simulations were run. The first run was performed using the unmodified version of \acidx{CLASS}. The second run used our amended version of \acidx{CLASS} with parameters according to Table~\ref{tab:planck2018iCDM}.
In what follows, we designate the cosmological constant by $\Lambda$, as customary. The effect of DE in the \newmodel{}, we denote by $\rho_{\aCC}$, $w_{\aCC}$ and $\Omega_{\aCC}$, respectively, for habit's sake; see Sec.~\ref{sec:EnhancedFriedmann}. \acidx{LCDM} refers to the concordance model, whereas the \newmodel{} is denoted as \q{wCDM}. 
This notion is often associated with the specific CPL parametrization \eqref{eq:EQCPLH0}, whereas we use it to refer to our own parametrization in \eqref{eq:EQwbackreactionCautun}. 
 
\subsection{Evolution of densities and equation of state}
\begin{figure} [!htbp]
	\includegraphics[width=1.0\columnwidth]{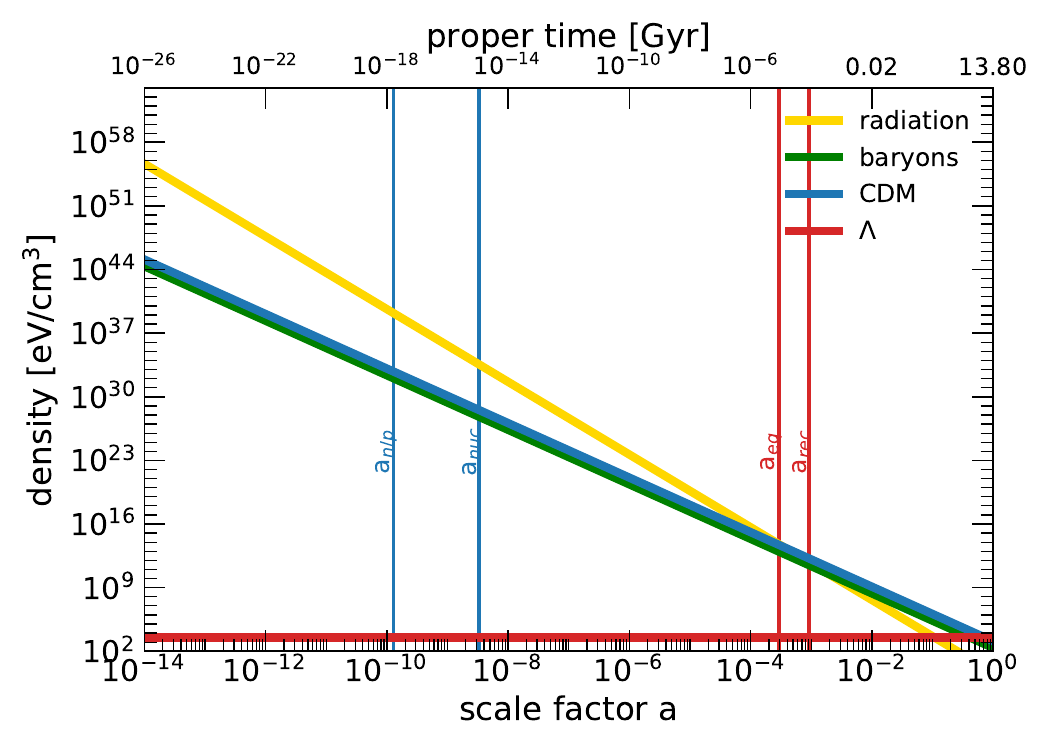}
	\includegraphics[width=1.0\columnwidth]{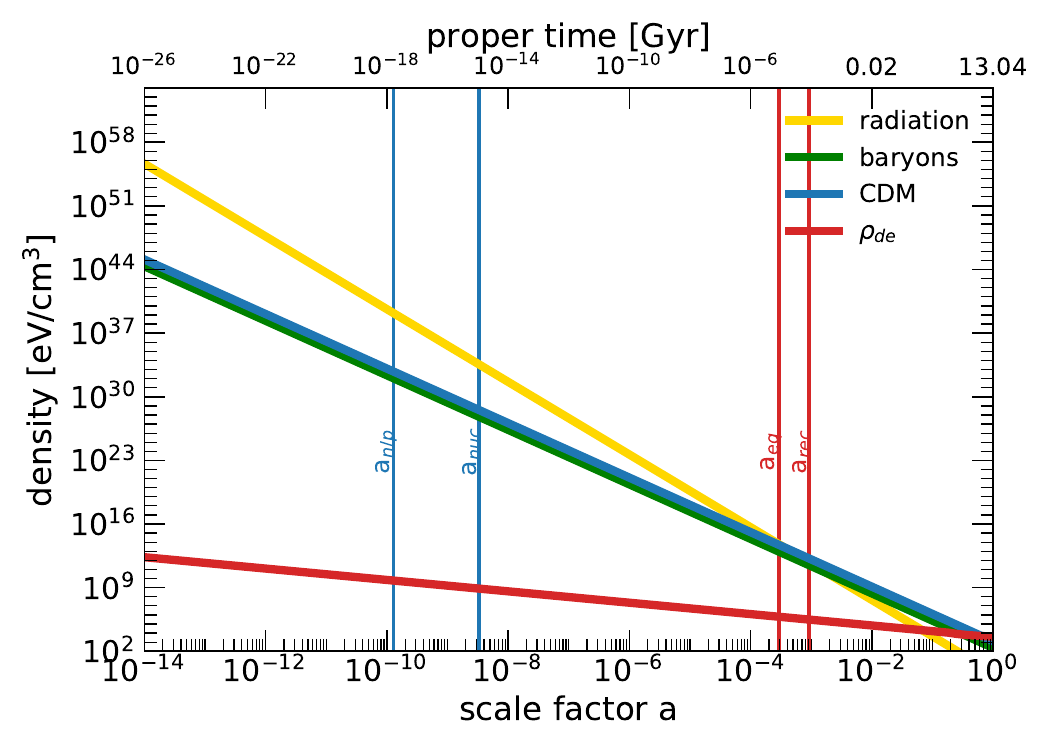}
	\caption[Evolution of densities in $\Lambda$CDM and our extension]
	{\tb{Evolution of energy densities in \boldmath{$\Lambda$CDM} (top panel) and \acidx{wCDM} (bottom panel) as a function of scale factor and proper time, respectively.} The top panel displays the evolution of the individual contributions to the energy density of the $\Lambda$CDM model according to \eqref{eq:EQdensitiesLCDMfull}. The red solid line indicates the evolution of the cosmological constant $\Lambda$. The bottom panel displays the evolution of the individual contributions to the energy densities of the \glo{wCDMmodel}. The radiation and matter components show no differences, compared to $\Lambda$CDM. However, the red solid line for $\rho_{\aCC}$ is not a constant, but has a slope, corresponding to $w_{\aCC,\sini} \sim -0.8$, according to \eqref{eq:EQfriedmannparamsSingle}. Note that the late impact of voids onto $w_{\aCC,\sini}$ is not recognizable in the log-log plot. The vertical lines in blue bracket the epoch of big bang nucleosynthesis, between neutron-proton freeze-out at $a_{\text{n/p}} \sim 1.3 \cdot 10 ^{-10}$ and nuclei production $a_\text{nuc} \sim 3.3 \cdot 10^{-9}$, while the vertical lines in red (at larger scale factor) indicate the time of matter-radiation equality $a_{\text{eq}}$, followed by recombination $a_{\text{rec}}$ (see also Fig.~\ref{fig:evol-Omega}).
	}
	\label{fig:evol-rho}
\end{figure}

%
Figure \ref{fig:evol-rho} displays the evolution of the energy densities $\rho_i$ vs scale factor $a$, as well as vs proper time, in both models.
%
%
The top panel displays the evolution of the densities in the \glo{LCDMmodel}, corresponding to the individual components of the Universe, as indicated in the legend. The slopes of the individual curves are as expected from Eqs.~\eqref{eq:EQdensitiesLCDMfull}. The red solid line depicts $\Lambda$, showing a constant evolution as expected for the cosmological constant. The bottom panel displays the evolution of the energy densities in the \glo{wCDMmodel}. In this model, the early slope of the red curve for $\rho_{\aCC}$ corresponds to the \acidx{EOS} parameter $w_{\aCC,\sini} \simeq -0.8$, determined by \eqref{eq:EQfriedmannparamsSingle} and \eqref{eq:EQwbackreactionCautun}. The vertical lines in blue bracket the epoch of big bang nucleosynthesis, between neutron-proton freeze-out at $a_\text{n/p} \sim 1.3 \times 10^{-10}$ and nuclei production at $a_\text{nuc} \sim 3.3 \times 10^{-9}$, while the vertical lines in red (at larger scale factor) indicate the time of matter-radiation equality $a_\text{eq} \sim 10^{-4}$, followed by recombination $a_\text{rec} \sim 10^{-3}$ (the detailed numbers for the latter depend somewhat more upon the cosmological model). The effect of the backreaction from voids in the late stages of the evolution is so small that it cannot be recognized in the log-log plot. The evolution of the matter and radiation components indicates no significant differences between the two models, as expected.\par
However, the ages of the models differ, as can be seen from the value of the proper time at the very present. The $\Lambda$CDM model has an age of $13.8$~Gyr, whereas the age of the \glo{wCDMmodel} is $13.04$~Gyr.  In Sec.~\ref{sec:hubbletension}, where we discuss the Hubble tension problem, we will present arguments to show that the lower age of the \acidx{wCDM} universe is no obstacle for the model.\par
\begin{figure} [!t]
	\includegraphics[width=1.0\columnwidth]{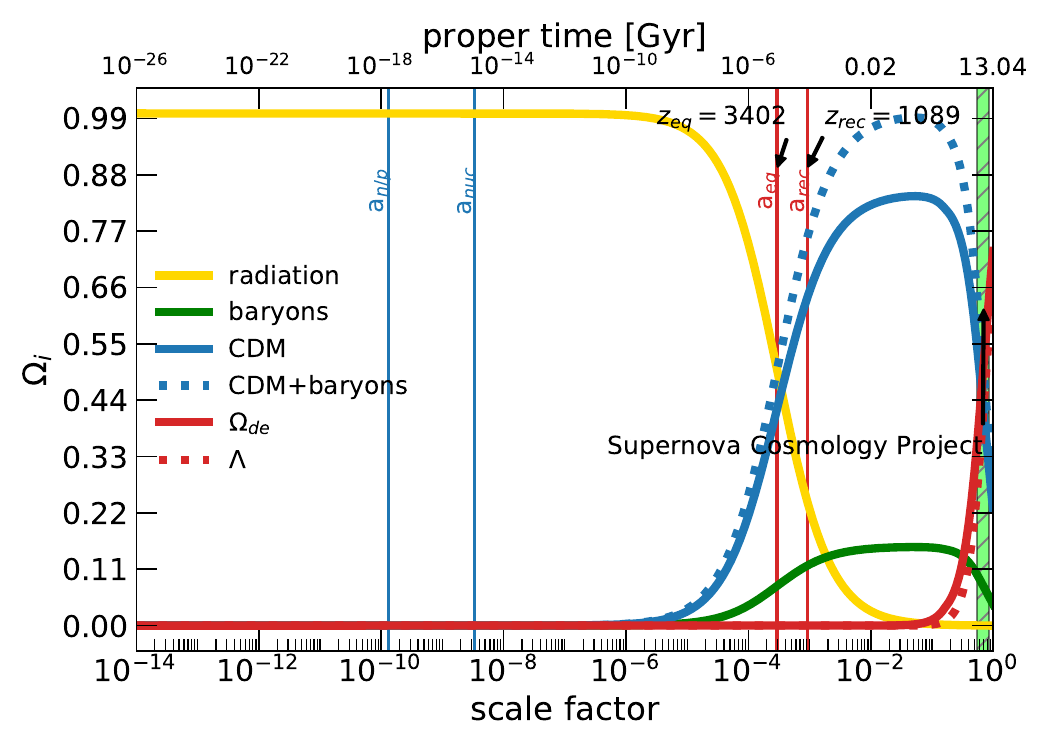}
	\includegraphics[width=1.0\columnwidth]{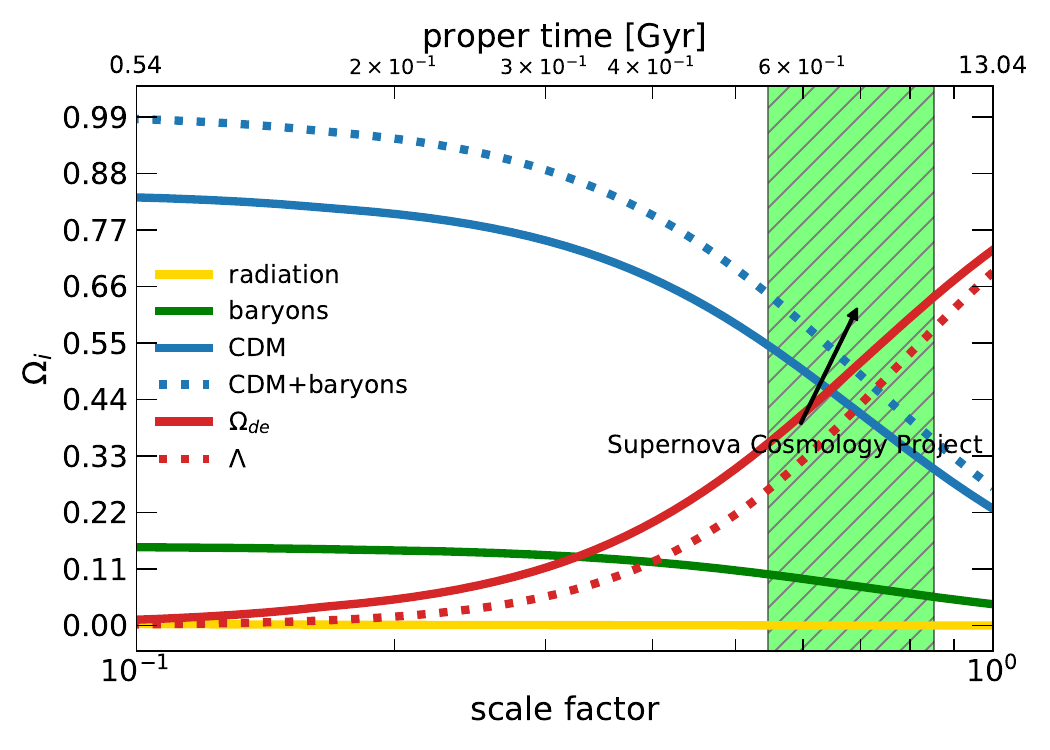}	
	\caption[Evolution of density parameters in \boldmath{$\Lambda$CDM} and our extension]
	{\tb{Evolution of density parameters in $\Lambda$CDM and \acidx{wCDM}.} The top panel displays the evolution of the density parameters in the \glo{wCDMmodel} vs scale factor and proper time, respectively. It includes also $\Omega_{\Lambda}$ of the $\Lambda$CDM model (red dotted curve) in order to compare it to $\Omega_{\aCC}$ (red solid curve) of the \glo{wCDMmodel}. Vertical lines are the same as in Fig.~\ref{fig:evol-rho}. The bottom panel displays the same quantities but shows a zoom-in to highlight the range of scale factors from $a=0.1$ to $1$. We see that $\Omega_{\aCC}$ of \acidx{wCDM} gets dominant earlier than $\Omega_{\Lambda}$ of $\Lambda$CDM, but that toward $a=1$ the two curves come closer again. As an illustration, the green-hatched vertical band indicates the range of scale factors of the sample of galaxies used by the Supernova Cosmology Project (\citet{Perlmutter1999}). More explanations can be found in the main text. 
	}
	\label{fig:evol-Omega}
\end{figure}
Figure \ref{fig:evol-Omega} displays the evolution of the density parameters $\Omega_i$ (as indicated in the legends) vs scale factor $a$ and proper time, respectively, for both models.
The panels display both $\Omega_{\aCC}$ of \acidx{wCDM} (red solid curve) and $\Omega_{\Lambda}$ of $\Lambda$CDM (red dotted curve). We can see that $\Omega_{\aCC}$ of \acidx{wCDM} gets to be the dominant contribution to the energy budget of the Universe at a smaller scale factor than $\Omega_{\Lambda}$ of $\Lambda$CDM. As an illustration, we include the green-hatched vertical band which indicates the range of redshifts (resp. scale factors) of the sample of galaxies used by the Supernova Cosmology Project (\citet{Perlmutter1999}). It was one of the high-redshift SNe Ia surveys that discovered the accelerated expansion in the late 1990s. The bottom panel displays the same quantities but shows a zoom-in to highlight the range of scale factors from $a = 0.1$ to $1$. 
While the advanced $\Omega_{\aCC}$ gets dominant earlier than the cosmological constant $\Lambda$, the two curves come close again at larger $a$, due to the backreaction from voids which makes $\rho_{\aCC}$ converge to a cosmological constant.\par
\begin{figure} [!b]
	\includegraphics[width=1.0\columnwidth]{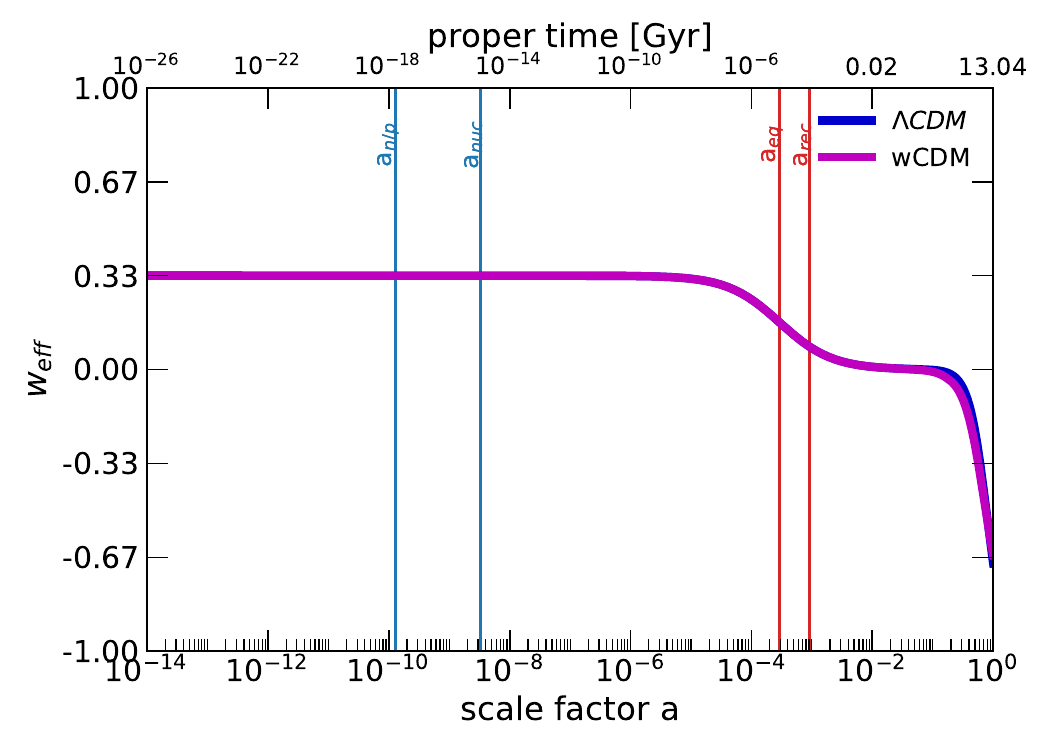}	
	\includegraphics[width=1.0\columnwidth]{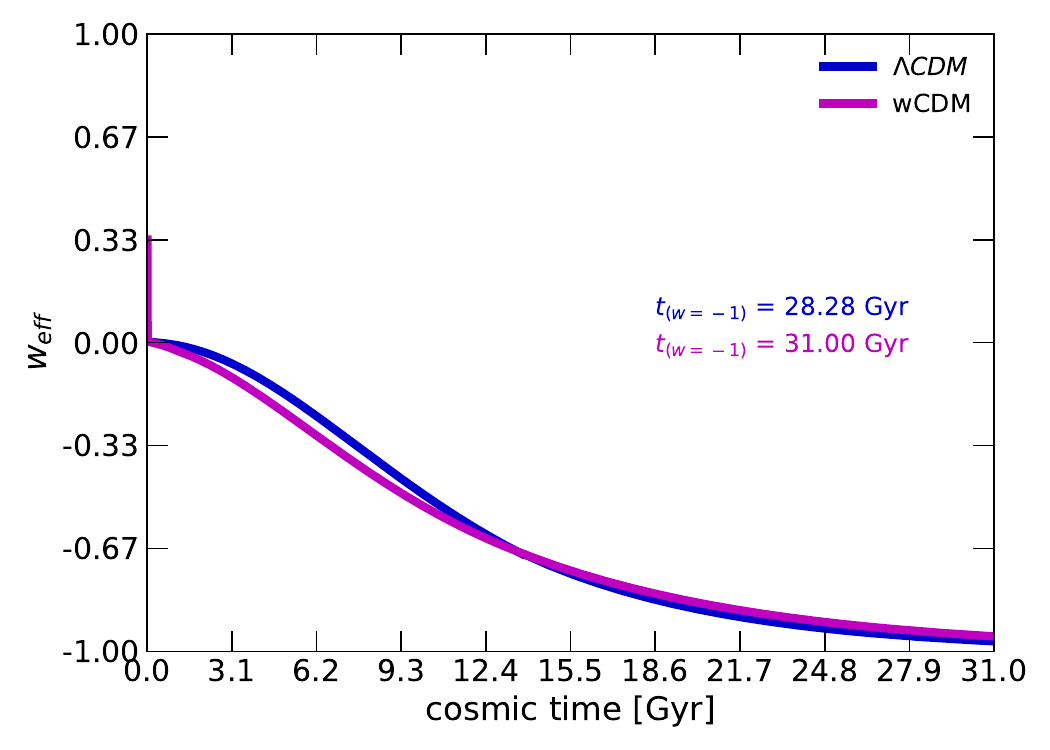}		
	\caption[Evolution of effective EoS parameter $w_{\text{eff}}$ in $\Lambda$CDM and our extension]
	{\tb{Evolution of effective EoS parameter \boldmath$w_\text{eff}$ in extended and concordance \boldmath$\Lambda$CDM}. The top panel displays the evolution of $w_\text{eff}$ in both models vs scale factor and proper time, respectively, with the same vertical lines as in Fig.~\ref{fig:evol-rho}. The bottom panel displays $w_\text{eff}$ vs cosmic time in linear scale! The cosmic time when $w_{\text{eff}} \simeq -1$ is indicated in the plot; it amounts to $ct = 28.28$ Gyr for $\Lambda$CDM vs $31.0$~Gyr for the \newmodel{}, respectively. 
	}
	\label{fig:evol-w}
\end{figure}
In Figure~\ref{fig:evol-w} we show the evolution of the effective \acidx{EOS} parameter for $\Lambda$CDM and our wCDM model, respectively. The \textit{effective} \acidx{EOS} parameter $w_\text{eff}(t) = p_{\text{tot}}(t)/\rho_{\text{tot}}(t)$ is determined by the time-dependent sum of pressures and densities of all cosmic components. (Generally, if one component dominates strongly over the others, $w_\text{eff}$ is close to the EoS parameter of that component.) 
In the top panel with the scale factor on the lower $x$-axis, we recognize the early radiation-dominated phase, where $w_\text{eff} = 1/3$ to high accuracy. In terms of cosmic time on a linear $x$-axis scale in the bottom panel, this phase is so short that it is seen as a vertical bar along the y-axis. Once matter strongly dominates in the Universe, shortly after $a_{\text{eq}}$, $w_\text{eff}$ drops to zero. In the latest phase, when $\Lambda$ begins to dominate the Universe in $\Lambda$CDM, the effective \acidx{EOS} parameter drops to its present-day value of $w_\text{eff} \sim -0.7$. On the other hand, in the \glo{wCDMmodel}, the effective \acidx{EOS} parameter initially decreases more rapidly, compared to $\Lambda$CDM. Only by the time the backreaction from voids comes at play in \acidx{wCDM}, the evolution approaches the one of $\Lambda$CDM, around the present, which can be also seen in the bottom panel of the figure.

In that bottom panel, we also highlight the evolution of the effective EoS parameter $w_\text{eff}$ in the far future for the two models. We can see that, as of a cosmic time of $t \sim 22$~Gyr, the $w_\text{eff}$ in both models, $\Lambda$CDM and \acidx{wCDM}, will show no significant difference, and $w_\text{eff}$ will be very close to $-1$ by the time of $t \sim 28$~Gyr for $\Lambda$CDM and $t \sim 31$~Gyr for \acidx{wCDM}.
However, while the dominance of $\Lambda$ as a cosmological constant is attributed to the former, it is the (relic) kinematic effect of the big bang via $H_{\text{ini}}$ in relation to the initial density, respectively, that is responsible for the dominance of $\rho_{\aCC}$ in the latter model. 

\subsection{Expansion history}\label{sec:iCDMhist}
\begin{figure} [!ht]
	\includegraphics[width=1.0\columnwidth]{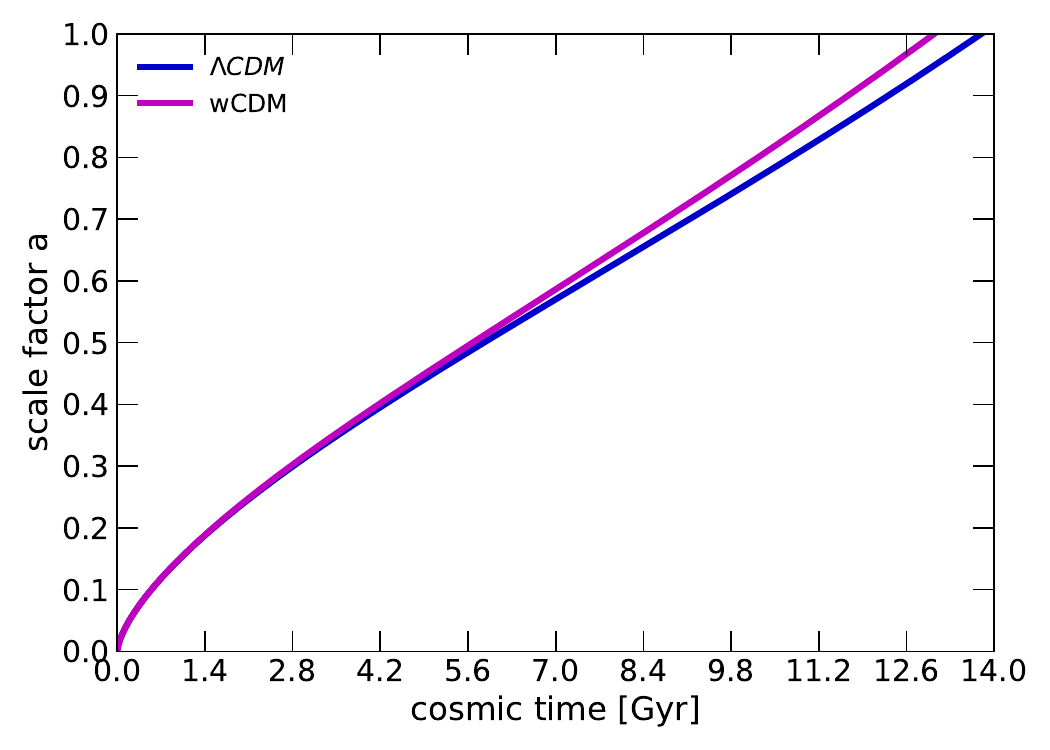}
	\includegraphics[width=1.0\columnwidth]{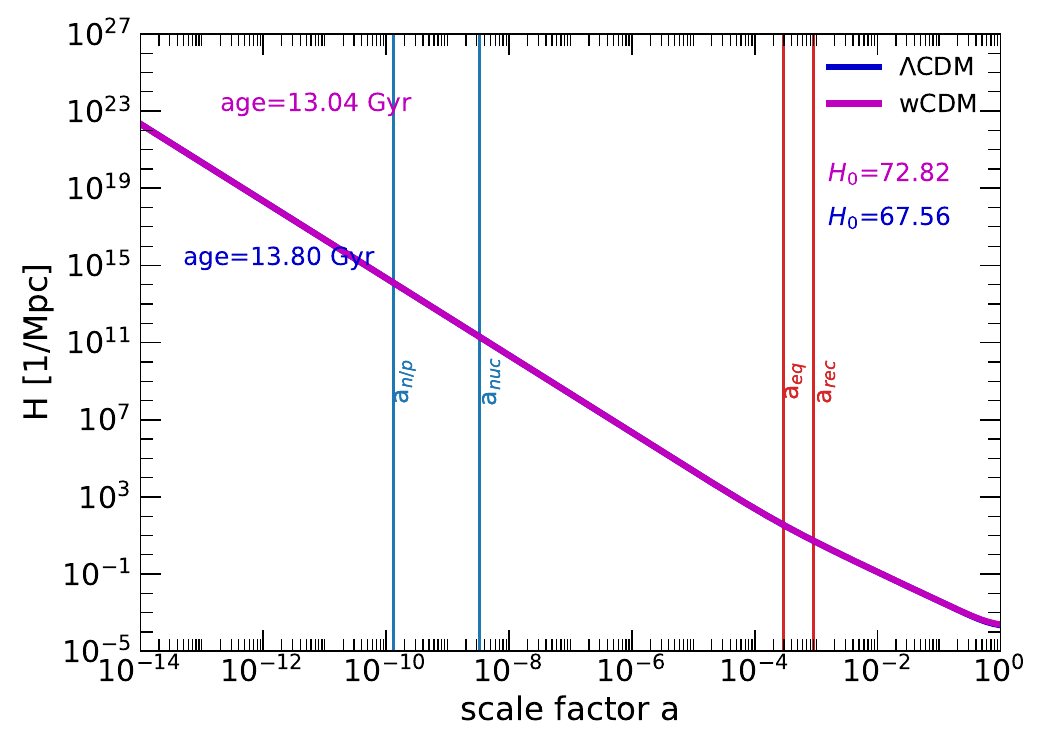}		
	\caption[Evolution of scale factor and Hubble parameter in $\Lambda$CDM and our extension]
	{\tb{Evolution of scale factor and Hubble parameter in $\Lambda$CDM and \acidx{wCDM}}. The top panel shows the evolution of the scale factor vs cosmic time for $\Lambda$CDM (blue curve) and \acidx{wCDM} (magenta curve); note the linear scales in this plot. The \glo{wCDMmodel} expands faster than the $\Lambda$CDM model, which can be better recognized in Fig.~\ref{fig:evol-hubble-diff}. The bottom panel shows the evolution of the expansion rate or Hubble parameter $H$ (displayed in \acidx{CLASS} conventions in units of $1/$Mpc) vs scale factor for both models. Its present-day value $H_0$ differs: $\Lambda$CDM has $H_0 = 67.56$ km/s/Mpc, whereas \acidx{wCDM} has $H_0 = 72.82$ km/s/Mpc. Vertical lines in the bottom panel are the same as in Fig.~\ref{fig:evol-rho}.
	}
	\label{fig:evol-hubble}
\end{figure}

%
Figure \ref{fig:evol-hubble} shows the expansion history of $\Lambda$CDM and \acidx{wCDM}.
The top panel displays the growth of the scale factor $a$ with cosmic time $t$ in linear scale for both models. The bottom panel shows the evolution of the Hubble parameter (i.e., expansion rate) $H$ vs scale factor $a$. In \acidx{wCDM} the scale factor grows faster than in $\Lambda$CDM. This results in a younger age of $13.04$~Gyr for \acidx{wCDM}, compared to $13.80$~Gyr for $\Lambda$CDM.\par
\begin{figure}  [!t]
	\includegraphics[width=1.0\columnwidth]{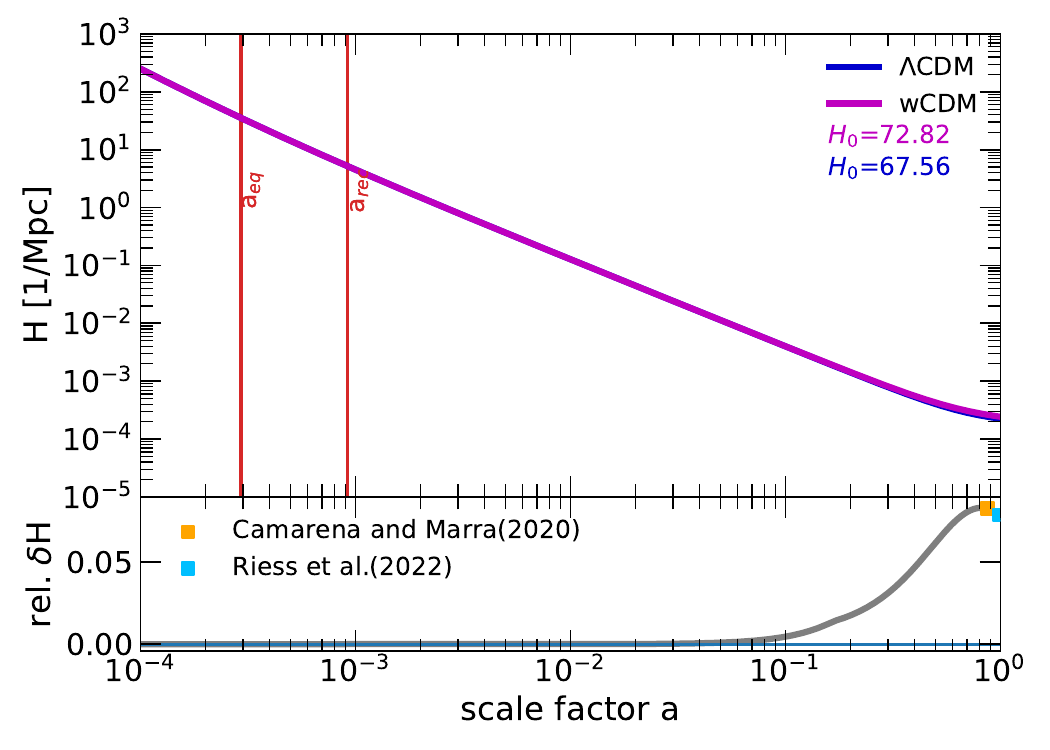}
	\includegraphics[width=1.0\columnwidth]{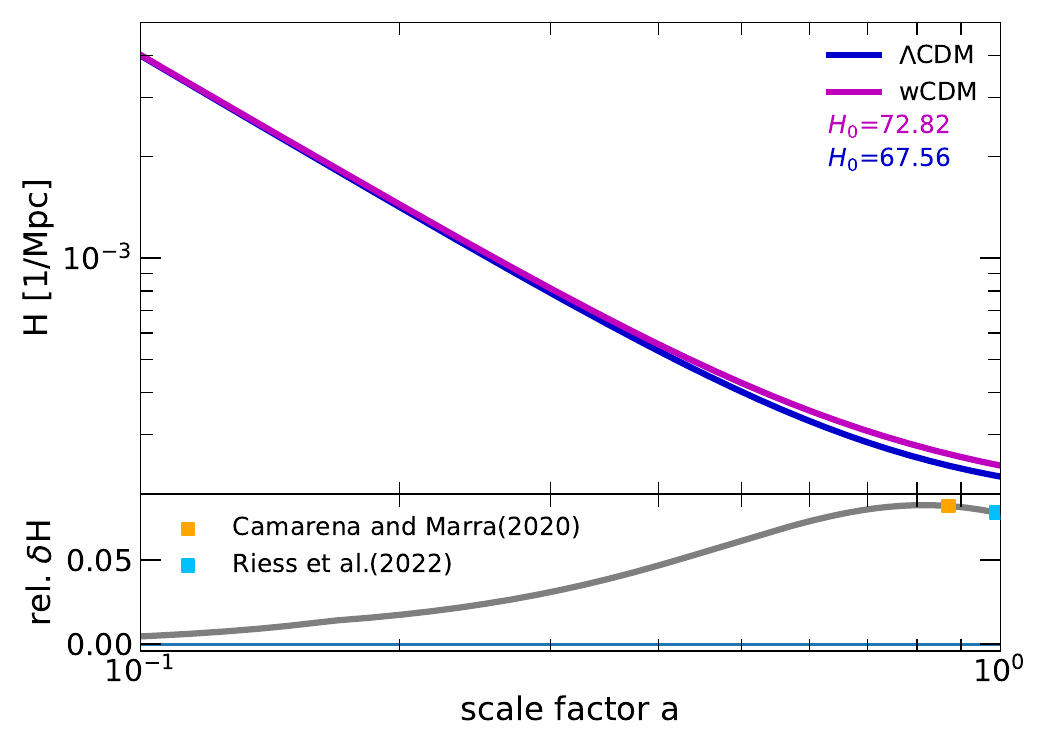}	
	\caption[Evolution of Hubble parameter vs scale factor]
	{\tb{Evolution of Hubble parameter vs scale factor}
		The top panel shows the evolution of the expansion rate $H$ vs scale factor for $\Lambda$CDM (blue curve) and the \newmodel{} (magenta curve) and their relative deviation as the gray curve in the bottom of each panel. The panels just differ in the range of scale factors shown. In the bottom panel, it can be recognized that the \newmodel{} expands faster than $\Lambda$CDM. The orange and the blue filled square indicate measurement values of $H_0$ by \citet{Camarena2020} and \citet{Riess2022}, respectively. The red vertical lines in the top panel are the same as in Fig.~\ref{fig:evol-rho}. In our \glo{wCDMmodel}, the Hubble parameter $H$ increases by $\sim 8$~\% by the present, compared to that of $\Lambda$CDM. As a result, the model provides an explanation for the Hubble tension problem and agrees with $H_0$-measurements at different redshifts.
	}
	\label{fig:evol-hubble-diff}
\end{figure}
The present-day value of the expansion rate, the Hubble constant $H_0$, also differs between the two models: $\Lambda$CDM has a value of $H_0 = 67.56$ km/s/Mpc and \acidx{wCDM} has $H_0 = 72.82$ km/s/Mpc. Of course, the value for $\Lambda$CDM is informed by \acidx{CMB} measurements from Planck, and is part of the parameters provided by \acidx{CLASS}, see Table \ref{tab:planck2018iCDM}.

Now, in order to illustrate the late impact by voids onto the expansion history of \acidx{wCDM}, compared to $\Lambda$CDM, Figure \ref{fig:evol-hubble-diff} shows zoom-ins of the evolution of the expansion rate $H$ for $a \geq 10^{-4}$ (top panel) and $a \geq 10^{-1}$ (bottom panel).
The vertical red lines indicate the scale factor at matter-radiation equality $a_{\text{eq}}$ and at recombination $a_\text{rec}$, respectively. The gray solid curve at the bottom of each panel displays the relative deviation of $H$ in \acidx{wCDM} from its value in $\Lambda$CDM. We can clearly see that at the time of recombination, when the CMB was emitted, there is no difference in $H$ between the models. However, at the late stages of the evolution, once the impact of the formation of the cosmic web becomes relevant, the expansion rate in the \glo{wCDMmodel} increases. At a scale factor of $a \sim 0.8$, the deviation reaches a maximum, but then decreases, yielding a value of $H_0 = 72.82$ km/s/Mpc. The bottom panel exemplifies this even more clearly. Again, this shift to a higher value of $H_0$ by $\approx 8$\% is due to the backreaction caused by voids.
The origin of the rise of $H_0$ in our \newmodel{} is the time-dependence (scale-factor dependence) of the \acidx{EOS} parameter $w_{\aCC}(a)$. As mentioned earlier, we cannot use the customary procedure applied in $\Lambda$CDM to determine the energy density $\rho_{\aCC}$ via \eqref{eq:EQdensevolgen}, as the \acidx{EOS} parameter $w_{\aCC}(a)$ is not a constant anymore. Hence, we have to perform a forward-in-time integration of $\rho_{\aCC}$ by integrating \eqref{eq:EQeconsnbg}, to solve the Friedmann equation \eqref{eq:EQfriedmannLCDMnbg}, which we discuss in detail in \citet{Foidl2024}. The consequence of this integration is that, as $w_{\aCC}(a)$ decreases with time, it leads to a rise in the value for $H_0$, and this value is higher compared to the parameterized Planck value given in Table~\ref{tab:planck2018iCDM}. As a result, the \newmodel{} can explain the Hubble tension problem, on which we elaborate below. 

\subsection{The Hubble Tension and \boldmath$\sigma_8$-Tension}\label{sec:hubbletension}
Our wCDM model of the previous section is phenomenologically similar to models of quintessence or \q{early dark energy} (EDE), which also propose solutions to the Hubble tension problem. A review on these models is given in \citet{Poulin2023}.
The Hubble tension problem refers to the reported discrepancy between the values\footnote{Strictly speaking, $H_0$ refers to $H$ at $z=0$, but since the community casually uses the notation $H_0$ also for any $H$ at low $z \lesssim 1$, we stick to this convention.} of $H_0$ derived from measurements of the CMB vs those derived from measurements in the \q{local} Universe, using standard candles such as SNe Ia or Cepheid variables. The measurements in the local Universe consistently reveal a higher value of $H_0$, than the measurements of the CMB. This problem is discussed for example in \citet{DiValentino2021}, where two measurement values in the local Universe are of particular interest to us for a comparison with our \newmodel{} (see their Fig. 1 and the bottom parts of both panels of our Figure~\ref{fig:evol-hubble-diff}): \citet{Riess2022} with $H_0 = 73.04$ km/s/Mpc and \citet{Camarena2020} with $H_0 = 75.4$ km/s/Mpc.
The \glo{wCDMmodel} with its value of $H_0 = 72.82$ km/s/Mpc agrees quite well with the value determined by \citet{Riess2022}. The result of \citet{Camarena2020} seems to be an outlier, as their $H_0$-value is even higher. The explanation for this can be found in the redshift range of the respective samples of galaxies for these two measurements. The redshifts of the sample used in \citet{Riess2022} lie between $z=0.011$ and $z=0.02$, whereas the sample in \citet{Camarena2020} is in a range between $z=0.023$ and $z=0.15$. Yet, because \acidx{wCDM} exemplifies a higher value of $H$ around that same epoch, it can also explain the results of \citet{Camarena2020}, whose measurement neatly lies on the maximum of the deviation of $H$ in the \glo{wCDMmodel}, compared to $\Lambda$CDM, see Figure ~\ref{fig:evol-hubble-diff}. This result came quite as a surprise when we studied this comparison in detail.

Now one of the drawbacks of these attempts is that mitigating the Hubble tension problem increases the $\sigma_8$ tension (see e.g. \citet{Kamionkowski2023}), which has emerged in recent years. This is a measure of the homogeneity of the Universe, where $S_8$ is defined as $\sigma_8 \sqrt{\Omega_m/0.3}$, with $\sigma_8$ being the standard deviation of the density fluctuation in a sphere of radius 8 $h^{-1}$~Mpc. A disagreement of more than 4$\sigma$ has been found between the extrapolation of the CMB temperature fluctuations forward to the present day, and what is measured by multiple probes of the inhomogeneity in the nearby Universe: The value of $S_8$ derived from the CMB is higher than that observed in the local Universe. A mitigation of the Hubble tension (i.e., by increasing $H_0$) increases the $\sigma_8$ tension.
In \citet{Reboucas2024}, this was addressed by assuming a modification of EDE in the late universe, which provides a solution to both problems. However, the modifications to EDE are purely empirically motivated, whereas our wCDM model is based upon a consistent set of physical processes involved in the evolution of the Universe.

In addition, we find that our 
$S_8 = 0.784$ almost exactly matches the value of the DES-Y3 measurements for which $S_8 = 0.782 \pm 0.019$, by assuming a constant neutrino mass $\sum M_{\nu}$. The final data release by the Planck mission in \citet{Tristram2024} also compares their result to this value. 
The result is in agreement with the findings by \citet{Reboucas2024}, who found that a time-dependent $w(z)$ at late times, can resolve Hubble tension \qemph{and} $\sigma_8$-tension. Indeed, this is the case in our extension \acidx{wCDM}, because the evolution of $w(z)$ in our model is constant in the early Universe, but in the late stages, it becomes a time-dependent function.

Now, in order to investigate whether the lower age of the \glo{wCDMmodel} constitutes a problem,
we compare some age indicators from the literature, notably estimates for the ages of the oldest known stars in the Universe. Their ages are estimated basically via two different methods: the abundance of heavy elements, mainly those formed by the r-process in core collapse supernovae; the second way is the determination of the time of the main sequence turn-off of metal-poor stars in globular clusters. A short review of methods and observations is given in \citet{Weinberg2008}, with age estimates of the oldest stars between $11.5$-$14$~Gyr, where the error bars are $\sim 2$~Gyr. The observations of metal-poor galaxies by \citet{Grebel2012} and  globular clusters in the local Universe by \citet{Grebel2016} display consistent results, albeit with larger error bars of $2-4$~Gyr. A more recent observation, based on the age of the metal-poor globular cluster NGC 6397 using WFC3/IR photometry is given in \citet{Correnti2018}, showing a similar age range with smaller error bars of $12.6 \pm 0.7$~Gyr. All these results are consistent with the age of our \acidx{wCDM} model.\par
For comparison's sake, we also computed the age of a $\Lambda$CDM universe using $H_0$ from \citet{Riess2022}, while keeping the other parameters unchanged: the result is $12.50$~Gyr. This age for $\Lambda$CDM would also be still on the save side with respect to the age estimates of the oldest known stars.

\citet{Cimatti2023} follow a different approach. They select the oldest objects from the literature and perform a statistical analysis within a $\Lambda$CDM cosmology, in order to determine the upper limit of $H_0$ compatible with the ages of the selected objects. They find $H_0 < 73.0 \pm 2.5$~km/s/Mpc with a probability of $93.2$~\%. So, the \q{Riess model} is just within these constraints. Since the expansion rate in \acidx{wCDM} increases only later in time, it is less affected by this upper limit.

Let us also comment on the early galaxies which have been found recently by the James Webb Space Telescope, see \citet{CurtisLake2023}, \citet{Finkelstein2023} and \citet{Robertson2023}. It has been estimated that these galaxies have formed only about 320 million years after the big bang. However, we stress that such estimates require the adoption of a specific cosmological model, in order to convert the measured redshifts into an age. Adopting a $\Lambda$CDM model with the $H_0$ determined by the Planck telescope, or the $H_0$ determined from the local Universe (e.g. by \citet{Riess2022}), will in each case shift the age of the galaxies accordingly, and the same is true if we adopt the \glo{wCDMmodel}, instead. In each case, there is no conflict between the age of these model universes and those early galaxies\footnote{Or, rephrased, all these cosmological models face a potential challenge in explaining the early formation of these galaxies.}.

Recent studies by e.g., \citet{Dainotti2022,Dainotti2021,Dainotti2022a,Bargiacchi2023,Bargiacchi2023a,OColgain2021} and \citet{Krishnan2021a}, indicate that the evolution of the expansion rate $H$ at low redshift is in conflict with the assumption of a cosmological constant, and dynamical models of DE seem to be preferred. These results may indicate that the origin of the Hubble tension is likely to be in cosmology and not in local measurement issues (e.g. \citet{Krishnan2021a}), as observations with increasing accuracy of $H_0$ in the local Universe seem to exacerbate the Hubble tension (e.g. \citet{Dainotti2023b} and \citet{RidaKhalife2024}). The evolution of the expansion rate, shown in Figure~\ref{fig:evol-hubble-diff}, displays a peak in the deviation to \acidx{LCDM} at $a \sim 0.8$. In fact, the position of this peak seems to be characteristic for individual DE candidates; see \citet{Foidl2024}. 

\section{Results for the owCDM Extension to \boldmath$\Lambda$CDM}\label{sec:resultsowCDM}

\begin{figure} [!htb]
	\includegraphics[width=1.0\columnwidth]{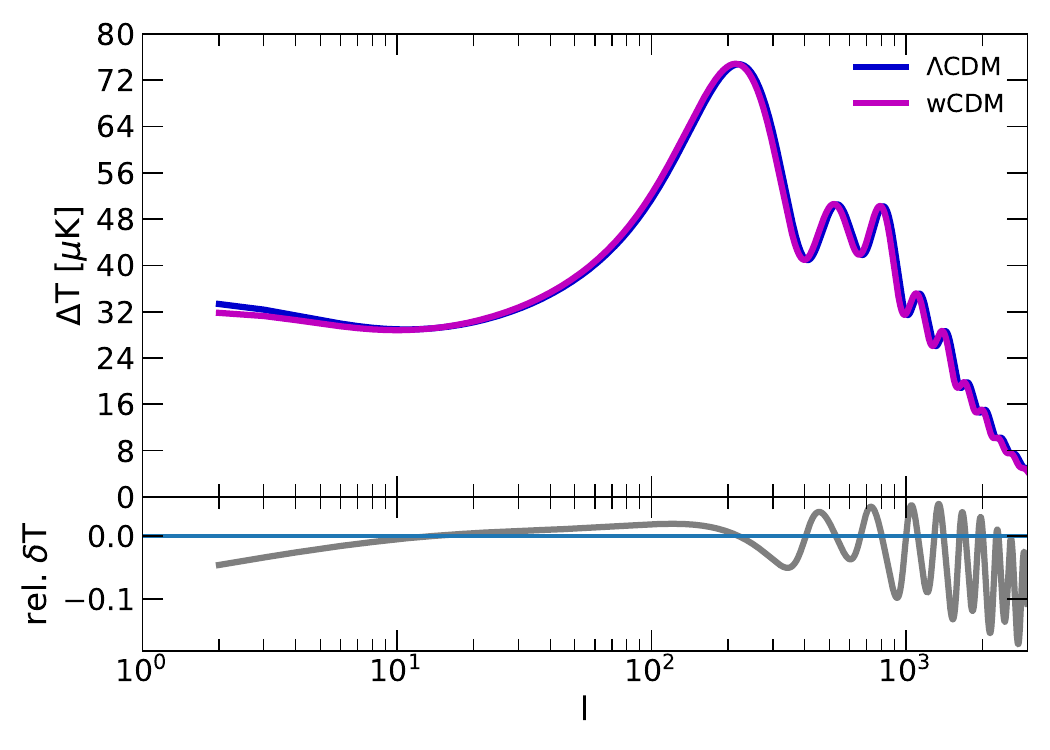}	
	\caption[The wCDM Temperature spectrum]
	{\tb{The CMB temperature power spectrum in $\Lambda$CDM (blue) and in our wCDM model (magenta)}. The relative differences are shown at the bottom of the panel, and amount to $\sim 15$ \%. 
	}
	\label{fig:evol-spectra}
\end{figure}

In the course of this section, we fit the \glo{wCDMmodel} to the \acidx{LCDM} \acidx{CMB} spectrum as well as to the value $\syidx{SYMH}_0 = 73.04$~km/s/Mpc, derived by \citet{Riess2022}, giving the \glo{owCDMmodel}.
In \acidx{CLASS}, we can readily incorporate the equations for \acidx{wCDM}, not only in the background module, but also in the perturbation module, in order to calculate perturbation spectra. Except for $\Omega_{\Lambda}$, \acidx{wCDM} uses the same cosmological parameters as concordance $\Lambda$CDM (see Table~\ref{tab:planck2018iCDM}). Also, it is based upon the standard CDM framework, concerning its matter content and the cosmic web (and void) structure. Therefore, we expect no big differences in the perturbation spectra between \acidx{wCDM} and concordance $\Lambda$CDM. In fact, the output by \acidx{CLASS} confirms this expectation, see Fig.~\ref{fig:evol-spectra}, where we show the power spectrum of the CMB temperature anisotropies of the \acidx{wCDM} and the \acidx{LCDM} model. 

Although we see no significant differences in the structure of the peaks in the CMB temperature power spectrum, the deviations are at a $\lesssim$ 15\% level. The structure of the peaks is almost identical, but the peaks are slightly shifted to the left, which can be explained as follows. The amplitudes of the peaks in the spectrum are determined by the density parameters $\Omega_{\gamma,0}$ for radiation, and $\Omega_{b,0}$ and $\Omega_{\text{CDM},0}$ for the matter components. The relative height of the peaks is determined by the ratio of $\Omega_{b,0}$ and $\Omega_{\text{CDM},0}$. Hence, it is obvious that the structure of the peaks is in agreement with \acidx{LCDM}, as we used the concordance values for these density parameters. However, the parameterization of \acidx{DE} impacts the expansion rate, as exemplified in Sec.~\ref{sec:iCDMhist}. Since there is no interaction between DE and the other cosmic components, the replacement of DE for $\Lambda$ does not impact the structure of the peaks. Keeping in mind that the multipole moments, shown on the x-axis, are in Fourier space, however, explains the shift of the spectrum to the left-hand side. More precisely, it is not a shift, but rather a stretching (compressing) of the spectrum from the right-hand side to the left-hand side, with increasing (decreasing) expansion rate. Thus, the slightly higher value of \Hconst{} of \acidx{wCDM} (see Fig.~\ref{fig:evol-hubble-diff}), compared to \acidx{LCDM} explains the minor shift to the left, while providing a solution to the Hubble tension. Remember that the cosmological parameter $\Omega_{k,0}$ for the spatial curvature has an analogous impact onto the spectrum. The degeneracy in the effect onto the spectrum between a change of spatial curvature and dark energy are well known, see e.g. \citet{Hu2002}.

In Sec.~\ref{sec:curvatureFLRW}, we argue that the free-falling, comoving \acidx{FLRW} observer resides in a local inertial system, where \acidx{SR} applies  and therefore perceives flat space.
In the concept of the \glo{flrwmetric}, the \acidx{FLRW} observer moves with freely streaming particles, which represent the galaxies in the Universe.
The observations of the \acidx{CMB} in space missions have determined, along with the flatness of space, the temperature fluctuations in the \acidx{CMB} with an accuracy of the order of $10^{-5}$~K around the average background temperature of roughly $2.7$~K. The expansion of the temperature \glo{powerspectrum} into spherical harmonics to display these measurements results do not include the first multipole moment, the dipole, for practical purposes. The dipole moment is induced by our peculiar motion against the \acidx{CMB}, and it is subtracted from the \acidx{CMB} raw data, because its amplitude of the order of $10^{-3}$~K would strongly dominate the other signal of interest. However, the fact that this dipole moment is still small, compared to the mean, just confirms that we -- as observers here -- fulfill the criterion of being comoving \acidx{FLRW} observers, not strictly (\citet{Peacock1999}), but to a very high degree (a well-known fact, but see also e.g., \citet{Hausegger2024}). Indeed, this observational finding of a small dipole actually confirms that we comove with the expansion to a high degree. So, the  metric we perceive is very close to the flat Minkowski metric.

\begin{figure} [!htb]
	\includegraphics[width=1.0\columnwidth]{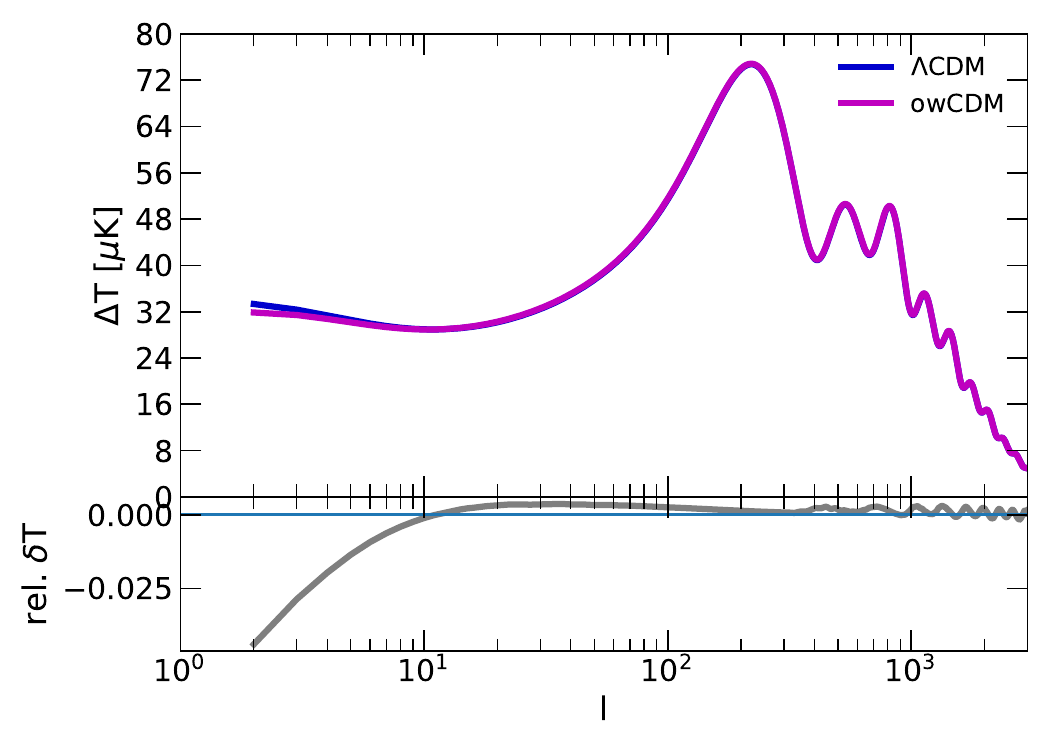}	
	\caption[The owCDM Temperature spectrum]
	{\tb{The CMB temperature power spectrum in $\Lambda$CDM (blue) and in our owCDM model (magenta).} The relative differences are shown at the bottom of the panel, and amount to $\sim 0.01$\%.
	}
	\label{fig:evol-spectra-fit}
\end{figure}

The dipole, which corresponds to our peculiar motion against the \acidx{CMB}, however, offsets us from the perfectly comoving \acidx{FLRW} observer, who observes a perfectly flat space.
Consequently, we now expand the scope and take into account our peculiar motion relative to the  \acidx{CMB}, caused by our \q{local} cosmological environment (see e.g., \citet{Tully2014}), considering the spatial curvature $\syidx{SYMdenspar}_k$ (since $\syidx{SYMdenspar}_k=0$ only applies to the perfectly comoving \acidx{FLRW} observer).
As is customary, we call the resulting model \glo{owCDMmodel}\footnote{This acronym is customarily used for extensions to \acidx{LCDM}, when $\syidx{SYMeos}(z)$ is different from $-1$ and additionally a spatial curvature is applied; see, for example, Particle Data Group Review Chapter 28 Dark Energy \cite{PDG202223}.}. The results are depicted in Fig. ~\ref{fig:evol-spectra-fit}.
Now, the deviation from the \acidx{LCDM} spectrum amounts to a $\lesssim$ $0.01$\% level (disregarding the higher deviations at small $l$, where the error bars in the actual \acidx{CMB} measurements are larger anyway).

We find a spatial curvature of $\Omega_{k,0} = -0.0197$ that describes a local effect, induced by our \q{local} attractive cosmic environment, decoupling us from perfect Hubble flow (see \citet{Tully2014}). This value is compatible with the final PR4 of the Planck mission (\citet{Tristram2024}) with $\Omega_{k,0} = - 0.012 \pm 0.010$.

In addition, we find for the owCDM model 
$S_8 = 0.798$, which is within $1\sigma$ with DES-Y3's result of $S_8= 0.782 \pm 0.019$. 
Upon our fitting, we can improve our
equation \eqref{eq:EQwbackreactionCautun} which describes the evolution of $w_{\text{de}}$, according to
\begin{equation} \label{eq:EQwbackreactionCautunFit}%
	\begin{aligned}
		w_{\aCC}(a) &= \begin{cases}
			w_{\aCC,\sini} \text{~~~when~} a \leq 1/6\\
			w_{\aCC,\sini} + \left[\frac{(1/a-1) - 5}{5}\right] (w_{\aCC,\sini} + 0.91)\end{cases}\\
		~&\text{~~~~~~~~~~~~~~~~~~when~}a > 1/6.
	\end{aligned}
\end{equation}
To reiterate, we stress that the evolution of $w_{\aCC}$ with cosmic time and the spatial curvature $\Omega_{k,0} = -0.0197$ are both effects of the nonlinear structure formation. The time-dependence of $w_{\aCC}$ is a global effect by the voids dominating the volume of the Universe, whereas $\Omega_{k,0} = -0.0197$ is a local effect by our local attractive cosmic environment, decoupling us from perfect Hubble flow. 

\section{Summary}\label{sec:iCDMsummarydiscussion}

We proposed two extensions to \acidx{LCDM}, wCDM and owCDM, which include the following novelties, on which we comment in the course of this section:
a) We take the equivalence principle and the concept of the comoving \acidx{FLRW} observer at face value and find that the geometry of the Universe as given by the \glo{flrwmetric} and the curvature of space are two individual quantities, where we identify the geometry with the \acidx{DE} component of our proposed extensions to \acidx{LCDM}.
b) In the \acidx{FLRW} formalism, the density parameters of subcritical and supercritical universes are also normalized to \glo{criticaldensity} by considering the suitable amount of spatial curvature. In contrast to this, we consider the consequence of the equivalence principle that irrespective of the geometry (open, closed or flat) of a universe, observers in the reference frame of the comoving FLRW observer, always perceive flat space. Thus, comoving (i.e., free-falling) observers in subcritical, supercritical or universes at critical density likewise perceive spatial flatness.
c) The FLRW formalism and therefore also \acidx{LCDM} consider the energy densities in the early Universe as the initial conditions determining the expansion history of the Universe. In contrast, we consider these initial energy densities in relation to the post \glo{bigbang} expansion rate as the initial conditions. The relation between these two quantities determines the geometry of the Universe as described by the curvature of the \glo{flrwmetric}.
d) Customarily, the averaging problem only considers the density of (linear and non-linear) perturbations and finds that there is no significant backreaction onto the evolution of the background universe. We additionally consider the volume of perturbations and find a mild backreaction, caused by voids dominating the volume of the Universe, providing a possible solution to the Hubble tension problem.
e) In owCDM we include the effect of the \acidx{CMB} dipole, which is otherwise not included in the calculations of \acidx{FLRW} models. This provides an explanation for the curvature of space measured by PR4 of the Planck mission (\citet{Tristram2024}) as the mere effect of our peculiar motion against the \acidx{CMB}, which \q{decouples} us from the Hubble flow.

We argued in Sec.~\ref{sec:curvatureFLRW} that the observation of flat space does not necessarily imply that we live in a universe with global flat space, expressed by zero curvature in the Einstein equations.
In fact, an observation of flat space, as perceived by the comoving \acidx{FLRW} observer in the local inertial system, on the grounds of the equivalence principle, should apply likewise to flat, open and closed geometries of the \glo{flrwmetric}.
In Sec.~\ref{sec:FLRWgeometry}, we exemplified that the initial conditions in the early Universe are given by the initial energy densities \qemph{and} the initial expansion rate obtained from the \glo{bigbang}. 
 
In order to account for arbitrary expansion rates obtained from the \glo{bigbang}, not just the fine-tuned case of \acidx{LCDM}, in Sec.~\ref{sec:EnhancedFriedmann} we incorporate the initial expansion rate into the \acidx{LCDM} formalism. We identify the geometry of the Universe induced by the \acidx{FLRW} metric and the curvature term in the \glo{friedmannequation} \eqref{eq:EQfriedmannsgeneralNLK}, respectively, with $\syidx{SYMed}_{\aCC}$ and express the perceived spatial flatness by $\syidx{SYMed}_{k}=0$. We associated an \acidx{EOS} with $\syidx{SYMed}_{\aCC}$, which depends on the amount of the energy content in the Universe (different from the operationally defined usual quantities $\syidx{SYMed}_{\aCC}$ and $\syidx{SYMed}_{k}$).
Thus, $\syidx{SYMed}_{\aCC}$ (along with $\syidx{SYMeos}_{\aCC}$) describes the dynamics of the expansion of the background universe, as determined by the \acidx{IC} of the Universe, and is not a \qc{physical substance}{,}  in contrast to the assumption of the \glo{cosmoconstant} \CCLambda{}. Nevertheless, we use the fluid formalism to include its impact into the energy momentum tensor $\syidx{SYMemt}_{\mu\nu}$, just in the same way it is done in \acidx{LCDM}.  
The \glo{wCDMmodel} explains the phenomenology of a late-time accelerated expansion as a kinematic effect, induced by the primeval expansion rate $\syidx{SYMH}_{\text{ini}}$ in relation to the initial density, originating from the \glo{bigbang}. 
By choosing the same present-day matter content of $\syidx{SYMdenspar}_{m,0} \simeq 0.3$ as in \acidx{LCDM}, we find $\syidx{SYMeos}_{\aCC} \simeq -0.8$, very close to $-1$ of a \glo{cosmoconstant}.
To reiterate, the Universe is, according to the \glo{wCDMmodel}, described as subcritical Universe. This is different to \acidx{LCDM}, where the \glo{cosmoconstant} \CCLambda{} is thought of as a kind of \q{physical substance} just like matter or radiation, as part of the energy-momentum tensor, which renders the Universe to be at \glo{criticaldensity}.

In the subsequent Sec.~\ref{sec:largescale}, we reconsidered the impact of the nonlinear structure formation in the late epochs of the Universe, especially in the form of its cosmic web, once voids dominate the Universe in volume and \mbox{(under-)density}.
We incorporated the impact of backreaction from cosmic voids onto the expansion history by way of its effect onto the \acidx{EOS} parameter $w_{\aCC}$. As a result, the initially constant $w_{\aCC}$ morphs into a time-dependent function. We derived this function, using void models informed by cosmological $\Lambda$CDM simulations of the previous literature. This is justified, since we adopt the same standard CDM paradigm in our $\Lambda$CDM extensions (wCDM and owCDM).
In Sec.~\ref{sec:resultsiCDM}, we compared the results from computations of \acidx{LCDM} and \acidx{wCDM}. The initially constant EoS parameter, $w_{\aCC} \simeq -0.8$, evolves to a more negative value of $w_{\aCC} \simeq -0.9$ at the present time, once the impact by voids has become significant. While this impact by voids onto the expansion history affects an overall change of only about $8\%$ in the expansion rate of the late Universe, it is enough in order for our wCDM model to explain the Hubble tension problem, which refers to the offset between measurements of the Hubble constant $H_0$, using the CMB versus standard candles in the \q{local} Universe. 
We implemented \acidx{wCDM} into our amended version of the open-source Boltzmann code \acidx{CLASS}, in order to calculate the expansion history and the linear power spectra of this model. We compared our results with concordance $\Lambda$CDM, as well as with observations, and found that \acidx{wCDM} agrees well with current data, such as the Dark Energy Survey. 

Owing to the different evolution of $\rho_{\aCC}$ in our \newmodel{} versus $\Lambda$ in $\Lambda$CDM, there is a difference in the age of the models, namely $13.04$ Gyr versus $13.8$ Gyr, respectively. We checked to see that the younger age of \acidx{wCDM} is not in conflict with age estimates of the oldest known stars. It is also not in conflict with the very early galaxies found by the James Webb Space Telescope, since the redshift determinations of these galaxies require a cosmological model to convert redshift into age. In a \acidx{wCDM} universe, these galaxies would be correspondingly younger than in a $\Lambda$CDM universe. 

Furthermore, there are indications from observations that the evolution of the expansion rate $H$ at low redshift is in conflict with the assumption of a cosmological constant, and dynamical models of DE with a time-dependent \acidx{EOS} parameter may be preferred; see for example \citet{Dainotti2022,Dainotti2021,Dainotti2022a,Bargiacchi2023,Bargiacchi2023a,OColgain2021} and \citet{Krishnan2021a}. These results may indicate that the origin of the Hubble tension is likely to be in cosmology and not in local measurement issues, which we discuss in detail in \citet{Foidl2024}. Also, in that paper we find that the Hubble tension problem can be explained phenomenologically in a very natural way by a DE component with a decreasing \acidx{EOS} parameter $w(z)$. Our $\Lambda$CDM extension described in this paper fulfills this property. 

In the asymptotic future, the EoS parameter $w_{\aCC}$ converges to the one of a cosmological constant with a value of $-1$ (see Fig.~\ref{fig:evol-w}). In fact, the future point in time where the effective EoS parameter (including all cosmic components) is very close to this value of $-1$ is not much different between \acidx{wCDM} and concordance $\Lambda$CDM.

In the final step, we compared the CMB temperature spectra of \acidx{wCDM} and \acidx{LCDM}. We find that \acidx{wCDM}'s spectrum is shifted slightly to the left-hand side, while the amplitudes and structure of the peaks match perfectly. We explain this shift as the effect due to the increase of the expansion rate in our model, compared to \acidx{LCDM}. Subsequently, we fitted \acidx{wCDM} to \acidx{LCDM}'s CMB temperature spectrum, as well as to the locally measured value of $\syidx{SYMH}_0 = 73.04$ by \citet{Riess2022}, while we additionally considered spatial curvature. The resulting owCDM model reproduces \acidx{LCDM}'s CMB spectrum at a 0.01\% level, where the model parameters additionally include  $\Omega_{k,0}=-0.0197$, which is very well compatible with the final PR4 of the Planck mission (\citet{Tristram2024}). We explain the spatial curvature as an effect of our \q{local} attractive cosmological environment (see e.g. \citet{Tully2014}), which offsets us from the position of a perfectly comoving FLRW observer, described by $\Omega_{k,0}=0$, thus decoupling us from the Hubble flow, seen as a dipole in the CMB measurements. This dipole does not enter the computations in \acidx{LCDM}, though. 
In addition to providing a solution to the Hubble tension, we find that both, wCDM and owCDM, also mitigate the $\sigma_8$ tension.

In conclusion, we like to note that we presented two extensions to \acidx{LCDM}, which are based on a time-dependent evolution of the \acidx{EOS} parameter $w(z)$ of the \q{\acidx{DE} component}. In contrast to other parameterisations, like the empirically defined CPL parameterisation, we conceived our \acidx{DE} model based on well-accepted concepts and theories, already applied in \acidx{LCDM}. Furthermore, our extensions wCDM and owCDM can be tested, using ongoing and future observational campaigns, within their current scope. Only minor adaptation to the process of data analysis is required to implement our extensions into these campaigns, as we also exemplify in \citet{Foidl2024}, and where we present a list of observational programs of interest.
By determining the \acidx{EOS} of \acidx{DE}, these programs will be able to rule out our extensions or confirm them by providing an even more accurate update of our function of $w_{\text{de}}(z)$, i.e. a more observationally informed EoS, compared to our approximations, based on the Millennium simulation, in this paper.
%
%
%
\begin{acknowledgements}
    The authors are grateful to Glenn van de Ven, Paul Shapiro, Dragan Huterer, Oliver Hahn and Bodo Ziegler for helpful and valuable discussions, concerning an earlier version of this manuscript.
    T.R.-D. acknowledges the support by the Austrian Science Fund FWF through the FWF Single-Investigator Grant (FWF-Einzelprojekt) No. P36331-N, and the support by the Wolfgang Pauli Institute in hosting this grant.  
\end{acknowledgements}
%
%

%
%
\end{document}